\documentclass[lettersize,journal]{IEEEtran}
\usepackage{amsmath,amsfonts}
\usepackage{algorithmic}
\usepackage{algorithm}
\usepackage{array}
\usepackage{textcomp}
\usepackage{stfloats}
\usepackage{url}
\usepackage{verbatim}
\usepackage{graphicx}
\usepackage{cite}
\usepackage{cases}
\usepackage{xcolor}

\newtheorem{theorem}{Theorem}
\newtheorem{corollary}{Corollary}
\newtheorem{lemma}{Lemma}
\newtheorem{definition}{Definition}

\newtheorem{proposition}{Proposition}

\usepackage{caption}
\usepackage{subfigure}

\newcommand*{\Scale}[2][4]{\scalebox{#1}{$#2$}}%
\begin{document}

\title{To Analyze and Regulate Human-in-the-loop Learning for Congestion Games}

\author{Hongbo Li,~\IEEEmembership{Member,~IEEE}, and Lingjie Duan,~\IEEEmembership{Senior Member,~IEEE}
\thanks{Hongbo Li and Lingjie Duan are with the Pillar of Engineering Systems and Design, Singapore University of Technology and Design, Singapore 487372 (e-mail: hongbo\_li@sutd.edu.sg; lingjie\_duan@sutd.edu.sg). Part of this work was published in AAAI 2023 Conference \cite{li2023when}.}
\thanks{This work is also supported in part by the Ministry of Education, Singapore, under its Academic Research Fund Tier 2 Grant with Award no. MOE-T2EP20121-0001; in part by SUTD Kickstarter Initiative (SKI) Grant with no. SKI 2021\_04\_07; and in part by the Joint SMU-SUTD Grant with no. 22-LKCSB-SMU-053.} 
}

\markboth{IEEE TRANSACTIONS ON NETWORKING}%
{Shell \MakeLowercase{\textit{et al.}}: A Sample Article Using IEEEtran.cls for IEEE Journals}

\IEEEpubid{0000--0000/00\$00.00~\copyright~2021 IEEE}

\maketitle

\begin{abstract}
In congestion games, selfish users behave myopically to crowd to the shortest paths, and the social planner designs mechanisms to regulate such selfish routing through information or payment incentives. However, such mechanism design requires the knowledge of time-varying traffic conditions and it is the users themselves to learn and report past road experiences to the social planner (e.g., Waze or Google Maps). When congestion games meet mobile crowdsourcing, it is critical to incentivize selfish users to explore non-shortest paths in the best exploitation-exploration trade-off. First, we consider a simple but fundamental parallel routing network with one deterministic path and multiple stochastic paths for users with an average arrival probability $\lambda$. We prove that the current myopic routing policy (widely used in Waze and Google Maps) misses both exploration (when strong hazard belief) and exploitation (when weak hazard belief) as compared to the social optimum. Due to the myopic policy's under-exploration, we prove that the caused price of anarchy (PoA) is larger than \(\frac{1}{1-\rho^{\frac{1}{\lambda}}}\), which can be arbitrarily large as discount factor \(\rho\rightarrow1\). To mitigate such huge efficiency loss, we propose a novel selective information disclosure (SID) mechanism: we only reveal the latest traffic information to users when they intend to over-explore stochastic paths upon arrival, while hiding such information when they want to under-explore. We prove that our mechanism successfully reduces PoA to be less than~\(2\). 
Besides the parallel routing network, we further extend our mechanism and PoA results to any linear path graphs with multiple intermediate nodes. 
In addition to the worst-case performance evaluation, we conduct extensive simulations with both synthetic and real transportation datasets to demonstrate the close-to-optimal average-case performance of our SID mechanism.
\end{abstract}


\section{Introduction}
\IEEEPARstart{I}{n} transportation networks with limited bandwidth, mobile users tend to make selfish routing decisions in order to minimize their own travel costs. Traditional congestion game literature studies such selfish routing to understand the efficiency loss using the concept of the price of anarchy (PoA) (e.g., \cite{roughgarden2002bad,cominetti2024ordinary,bilo2020price,hao2022price}). To regulate the selfish routing behavior of atomic or non-atomic users and reduce social costs, various incentive mechanisms have been designed, including monetary payments to penalize users traveling on undesired paths (e.g., \cite{brown2017optimal,ferguson2021effectiveness,li2022online}). In practice, it may be difficult to implement such complicated payments and billing on users (\!\! \cite{pilz2013does,randhawa2019crowdsourcing}). This motivates the design of non-monetary mechanisms such as information restriction to influence selfish users to change their routing decisions towards the social optimum (e.g., \cite{tavafoghi2017informational,sekar2019uncertainty,castiglioni2021signaling}).
However, these works largely assume that the social planner has full information on all traffic conditions, and only consider one-shot static scenarios. This limits their practicality and applicability in real-world scenarios where information is incomplete for the social planner and dynamically changes over time.

In common practice, it is difficult to predict time-varying traffic conditions in advance (\!\!\cite{nikolova2011stochastic}). To obtain such information, emerging traffic navigation platforms (e.g., Waze and Google Maps) crowdsource mobile users to learn and share their observed traffic conditions on the way (\!\!\cite{vasserman2015implementing} and \cite{zhang2018distributed}). 
Nevertheless, these platforms simply expose all collected information to users as a public good. Consequently, current users often opt for selfish routing decisions, favoring paths with the shortest travel times instead of diversifying their choices to gather valuable information on the other paths for future users. Given that the traffic conditions on stochastic paths alternate between different cost states over time, users in these platforms might miss enough exploration of different paths to reduce the future social cost.

There are some recent works studying information sharing among users in a dynamic scenario (\!\!\cite{meigs2017learning,wu2019learning,kumar2025towards,vu2021fast}). For example, \cite{meigs2017learning} and \cite{wu2019learning} make use of former users' observations to help learn the future travel latency and converge to the Wardrop Equilibrium under full information. Similarly, \cite{vu2021fast} designs an adaptive information learning framework to accelerate convergence rates to Wardrop equilibrium for stochastic congestion games. 
However, these works cater to users' selfish interests and do not consider any mechanism design to motivate users to reach the social optimum. To study the social cost minimization, multi-armed bandit (MAB) problems are also formulated to derive the optimal exploitation-exploration policy among multiple stochastic arms (paths) (\!\!\!\cite{gittins2011multi,krishnasamy2021learning,bozorgchenani2021computation}). For example, \cite{bozorgchenani2021computation} applies MAB models to predict network congestion in a fast-changing vehicular environment. However, all of these MAB works strongly assume that users upon arrival always follow the social planner’s recommendations and overlook users’ deviation to selfish routing. 

\IEEEpubidadjcol

When congestion games meet mobile crowdsourcing, how to analyze and incentivize selfish users to listen to the social planner's optimal recommendations is our key question in this paper. As traffic navigation platforms seldom charge users (\!\!\cite{wang2022dynamic}), we target at non-monetary mechanism design which nicely satisfies budget balance property in nature. Yet we cannot borrow those information mechanisms from the literature in mobile crowdsourcing, as they considered that traffic information is exogenous and does not depend on users' routing decisions (e.g., \cite{kremer2014implementing,papanastasiou2018crowdsourcing,li2017dynamic,li2019recommending}). 
For example, \cite{li2019recommending} considers a simple two-path transportation network, one with deterministic travel cost and the other alternates over time between a high and a low constant cost state due to external weather conditions. 
In their finding, a selfish user is always found to under-explore the stochastic path to learn the latest information there for future users. In our congestion problem, however, a user will add himself to the traffic flow and change the congestion information in the loop. Thus, one may imagine that users may not only under-explore but also over-explore stochastic paths over time.
Furthermore, since the congestion information (though random) depends on users' routing decisions, it is easier for a user to reverse-engineer the system states based on the platform's optimal recommendation. In consequence, the prior information hiding mechanisms (in \cite{tavafoghi2017informational,li2019recommending,zhu2022information}) become no longer efficient. 

There are two related papers that study the regulation of selfish users in congestion games by dividing multiple user arrivals into two groups and providing them with different informational incentives simultaneously (\!\!\cite{li2024distributed,li2024human}). However, the analytical focus of these papers differs significantly from this work. In \cite{li2024distributed} and \cite{li2024human}, the simultaneous arrival of multiple users naturally leads to a reduction in congestion as users tend to select different paths, resulting in minimal disparity between the myopic policy and the socially optimal policy. 
While our work considers a scenario where the myopic policy can cause zero exploration of stochastic paths, leading to arbitrarily large efficiency losses compared to the social optimum. Given this worse performance under the myopic policy, the mechanisms proposed in the aforementioned works are not applicable to the single-user arrival scenario analyzed in this work.

We summarize our key novelty and main contributions in this paper as follows.
\begin{itemize}
    \item \emph{Novel human-in-the-loop learning for congestion games:} To our best knowledge, this paper is the first to analyze and regulate atomic users' routing over time to reach the best exploitation-exploration trade-off by providing incentives. In Section \ref{section2}, we first model a dynamic congestion game in a transportation network of one deterministic path and multiple stochastic paths to learn by randomly arriving users themselves. When congestion games meet mobile crowdsourcing, our study extends the traditional congestion games (e.g., \cite{tavafoghi2017informational,sekar2019uncertainty,castiglioni2021signaling,zhu2022information}) fundamentally to create positive externalities of information learning benefits generated by users themselves.
    \item \emph{POMDP formulation and PoA analysis:} In Section \ref{section3}, we formulate users' dynamic routing problems using the partially observable Markov decision process (POMDP) according to hazard beliefs of risky paths. 
    Then in Section~\ref{section4}, we analyze both myopic routing policy (widely used by Waze and Google Maps) and socially optimal policy to learn stochastic paths' states, and prove that the myopic policy misses both exploration (when strong hazard belief) and exploitation (when weak hazard belief) as compared to the social optimum. Accordingly, we prove that the resultant price of anarchy (PoA) is larger than $\frac{1}{1-\rho^{\frac{1}{\lambda}}}$, which can be arbitrarily large as discount factor $\rho\rightarrow1$.
    \item \emph{Selective information disclosure (SID) mechanism to remedy efficiency loss:} In Section \ref{section5}, we first prove that the prior information hiding mechanism in congestion games makes PoA infinite in our problem. Alternatively, we propose a selective information disclosure mechanism: we only reveal the latest traffic information to users when they over-explore stochastic paths, while hiding such information when they under-explore. 
    We prove that our mechanism reduces PoA to be less than $\frac{1}{1-\frac{\rho^{\frac{1}{\lambda}}}{2}}$, which is no larger than $2$. Besides the worst-case performance, we further show our mechanism's close-to-optimal average-case performance with respect to multiple variables by using extensive simulations.
    \item \emph{Extensions to any linear path graph and time-varying distributions of stochastic paths.} In Section~\ref{section6}, we further extend our system model to encompass any linear path graph with multiple intermediate nodes, and allow the traffic status of each stochastic path evolves according to dynamic Markov chains. In this extended model, our analysis reveals that the PoA for the myopic policy is still greater than $\frac{1-\sigma\rho^{\frac{1}{\lambda}}}{1-\rho^{\frac{1}{\lambda}}}$. This PoA decreases with the maximum variation $\sigma$ of transition probabilities in the dynamic Markov chain and can still be arbitrarily large as $\rho \rightarrow 1$ and $\sigma\rightarrow 0$.
    And our SID mechanism still works efficiently to reduce PoA to less than $2$, regardless of the maximum variation $\sigma$. Perhaps surprisingly, we show that the average performance of our SID mechanism may improve from an increasing variation $\sigma$.
    Finally, in Section~\ref{section7}, we use real datasets to show the close-to-optimal average-case performance of our SID mechanism for a more general hybrid network than linear path graphs.
\end{itemize}



\begin{figure}[t]
    \centering
    \captionsetup{font={footnotesize}}
    \subfigure[A typical parallel transportation network with $N+1$ paths.]{\label{fig:congestion_game}\includegraphics[width=0.43\textwidth]{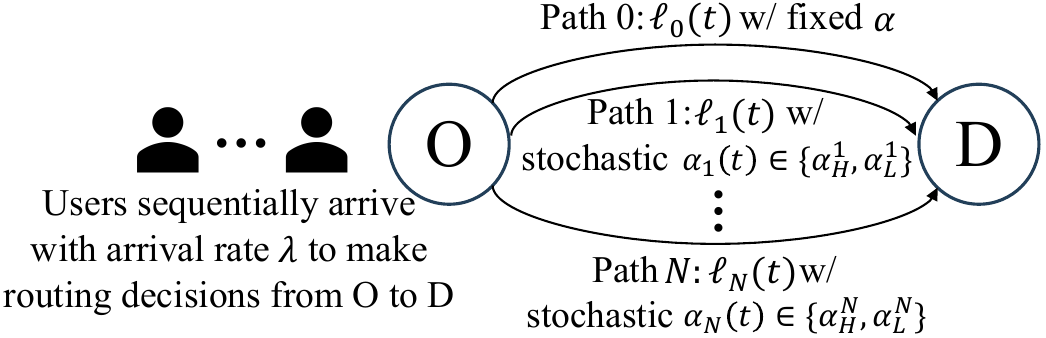}} 
    \subfigure[The partially observable Markov chain for modelling $\alpha_i(t)$ dynamics of stochastic path $i\in\{1,\cdots,N\}$.]{\label{fig:POMDP}\includegraphics[width=0.43\textwidth]{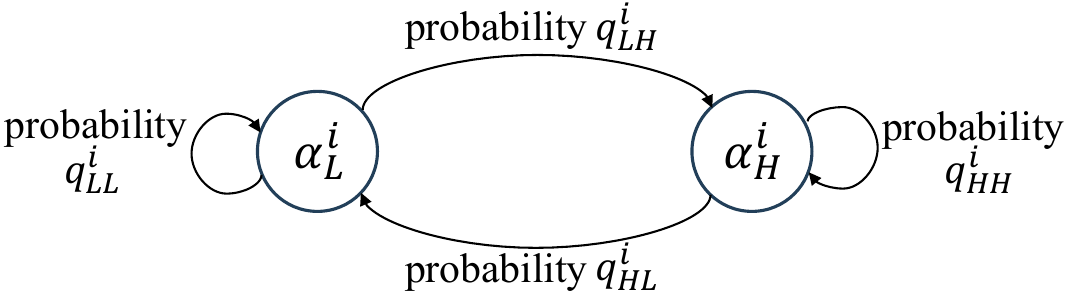}} 
    \caption{At the beginning of each time slot $t\in\{1,2,\cdots\}$, a user arrives with an average arrival probability $\lambda$ to choose a path among $N+1$ paths in the transportation network in Fig. \ref{fig:congestion_game}. The current travel latency $\ell_i(t)$ of each path $i\in\{0,1 ...,N\}$ has linear correlation with last latency $\ell_i(t-1)$ and evolves according to current user choice in (\ref{L_0(t+1)}) and (\ref{L_i(t+1)}). Path 0 is a safe route and its latency has a fixed correlation coefficient $\alpha\in(0,1)$ to change from the last round. Yet any risky path $i\in\{1,\cdots,N\}$ has a stochastic correlation coefficient $\alpha_i(t)$, which alternates between low coefficient state $\alpha_L^i\in[0,1)$ and high state $\alpha^i_H\geq 1$ according to the partially observable Markov chain in Fig. \ref{fig:POMDP}.}
    \label{pricing_fig}
\end{figure}

\section{System Model}\label{section2}
As illustrated in Fig. \ref{fig:congestion_game}, we consider a dynamic congestion game lasting for an infinite discrete time horizon. At the beginning of each time epoch $t\in\{1, 2, \cdots\}$, an atomic user arrives with an average arrival probability $\lambda$ to travel on one out of $N+1$ paths from origin O to destination D.\footnote{Here each time slot’s duration is properly selected to be short such that it is almost sure to have at most one user arrival at a time.}
Similar to the existing literature of congestion games (e.g., \cite{kremer2014implementing,tavafoghi2017informational,li2019recommending}), in Fig. \ref{fig:congestion_game} the top path 0 as a safe route has a fixed traffic condition $\alpha$ that is known to the public, while the other $N$ bottom paths are risky/stochastic to alternate between traffic conditions $\alpha_L^i$ and $\alpha_H^i$ over time, where the superscript $i$ represents the $i$-th risky path. Thus, the crowdsourcing platform expects users to travel to risky paths from time to time to learn the actual traffic information and plan better routing advisories for future users.

In the following, we first introduce the dynamic congestion model for the transportation network, and then introduce the users' information learning and sharing in the crowdsourcing platform. 
In the following, we first introduce the dynamic congestion model for the transportation network, and then introduce the users' information learning and sharing in the crowdsourcing platform. We summarize all the key notations of our paper in Table \ref{notation_table}.

\subsection{Dynamic Congestion Model}

\begin{table}[!t]
\renewcommand{\arraystretch}{1.3}
\caption{Key notations and their meanings in the paper}
\label{notation_table}
\centering
\begin{tabular}{|c|m{0.33\textwidth}|}
\hline
\textbf{Notation} & \textbf{Meaning}\\
\hline
\hline
$\lambda$ & The arrival probability of a user at each time slot.\\
\hline
$\ell_i(t)$ & Travel latency of path $i$ at time $t$.\\
\hline
$\pi(t)$ & Routing policy of the current user.\\
\hline
$s(t)$ & User arrival information in origin O at time $t$.\\
\hline
$\alpha, \alpha_i(t)$ & Correlation coefficients of safe path 0, risky path $i$.\\
\hline
$\alpha_H^i,\alpha_L^i$ & High, low hazard state on risky path $i$.\\
\hline
$\Delta \ell$ & Addition travel latency induced by each user.\\
\hline
$p_H,p_L$ & Probabilities for a user to observe a hazard under $\alpha_H,\alpha_L$. \\
\hline
$q^i_{HH},q^i_{LL}$ & Transition probabilities for the static Markov chain of risky path $i$. \\
\hline
$x_i(t)$ & Hazard belief of risky path $i$ at current time.\\
\hline
$\mathbf{L}(t)$ & Expected travel latency set of all $N+1$ paths.\\
\hline
$\mathbf{x}(t)$ & Hazard belief set of $N$ risky paths.\\
\hline
$\hat{\iota}(t)$ & The best risky path with the shortest expected travel latency among all $N$ risky paths.\\
\hline
$C^{(m)}(\cdot), C^*(\cdot)$ & Long-term cost functions under the myopic, socially optimal policy.\\
\hline
$\Bar{x}$ & Stationary hazard belief.\\
\hline
$\gamma$ & Average inefficiency ratio.\\
\hline
$\rho$ & Discount factor.\\
\hline
$q_{HH}^i(t),q_{LL}^i(t)$ & Stochastic transition probabilities for the dynamic Markov chain of risky path $i$.\\
\hline
$q_H,q_L$ & Expectations of $q_{HH}(t)$ and $q_{LL}(t)$.\\
\hline
$\sigma$ & Maximum variation of $q_{HH}(t)$ and $q_{LL}(t)$ from their expectations $q_{H}$ and $q_{L}$.\\
\hline
\end{tabular}
\end{table}

Let $\ell_i(t)$ denote the travel latency of path $i\in \{0,1,\cdots,N\}$ estimated by a new user arrival on path $i$ at the beginning of each time slot $t\in\{1,2,\cdots\}$. Define a binary variable $s(t)\in\{0,1\}$ to tell the user's arrival information at the origin O at the beginning of time $t$:
\begin{align*}
    s(t)=\begin{cases}
        1,&\text{if a user arrives with probability $\lambda$ at time }t,\\
        0, &\text{otherwise.}
    \end{cases}
\end{align*}
If there is a new user arriving at the origin O with $s(t)=1$, then this user decides the best path $i\in\{0,1,\cdots,N\}$ to choose by comparing the travel latencies among all paths. We denote a user's routing choice at time $t$ as $\pi(t)\in \{0,1,\cdots,N\}$. 
For this user, he predicts $\ell_i(t)$ based on the latest latency $\ell_i(t-1)$ and the last user's decision $\pi(t-1)$. If there is no user arrival with $s(t)=0$, then the current routing choice $\pi(t)=\emptyset$.

Some existing literature of delay pattern estimation (e.g., \cite{ban2009delay} and \cite{alam2019prediction}) assumes that $\ell_i(t+1)$ is linearly dependent on $\ell_i(t)$. Thus, for safe path 0 with the fixed traffic condition, its next travel latency $\ell_0(t+1)$ changes from $\ell_0(t)$ with constant correlation coefficient $\alpha$. Here $\alpha\in(0,1)$ measures the leftover flow to be serviced over time. Yet, if a new atomic user arrives and he chooses this path (i.e., $\pi(t)=0$ under $s(t)=1$), he will introduce an additional $\Delta \ell$ to the next travel latency $\ell_0(t+1)$, i.e., 
\begin{equation}
    \ell_0(t+1)=\begin{cases}
    \alpha \ell_0(t)+\Delta \ell,&\text{if }\pi(t)=0,\\
    \alpha \ell_0(t),&\text{if }\pi(t)\neq 0.
    \end{cases}\label{L_0(t+1)}
\end{equation}
Differently, on any risky path $i\in\{1,\cdots,N\}$, its correlation coefficient $\alpha_i(t)$ in this round is stochastic due to the random traffic condition (e.g., accident and weather change) at each time slot $t$. Similar to the congestion game literature \cite{meigs2017learning}, we suppose $\alpha_i(t)$ alternates between low coefficient state $\alpha_L^i\in[0,1)$ and high state $\alpha_H^i\in[1,+\infty)$ below:
\begin{equation*}
    \alpha_i(t)=\begin{cases}
    \alpha_L^i, &\text{if path } i\text{ has a good traffic condition at $t$,}\\
    \alpha_H^i, &\text{if path } i\text{ has a bad traffic condition at $t$.}
    \end{cases}
\end{equation*}
Note that we consider $\alpha_L^i<\alpha<\alpha_H^i$ such that each path can be chosen by users and we also allow jamming on risky paths with $\alpha_H^i\geq 1$. The transition of $\alpha_i(t)$ over time is modeled as the partially observable Markov chain in Fig.~\ref{fig:POMDP}, where the self-transition probabilities are $q_{LL}^i$ and $q_{HH}^i$ with $q_{LL}^i+q_{LH}^i=1$ and $q_{HH}^i+q_{HL}^i=1$. Then the travel latency $\ell_i(t+1)$ of any risky path $i\in\{1,\cdots,N\}$ is estimated as
\begin{equation}
    \ell_i(t+1)=\begin{cases}
    \alpha_i(t) \ell_i(t)+\Delta \ell,&\text{if }\pi(t)=i,\\
    \alpha_i(t) \ell_i(t),&\text{if }\pi(t)\neq i.
    \end{cases}\label{L_i(t+1)}
\end{equation}
To obtain this $\alpha_i(t)$ realization for better estimating future $\ell_i(t+1)$ in (\ref{L_i(t+1)}), the platform may expect the current user arrival to travel on this risky path $i$ to learn and share his observation.

\subsection{Crowdsourcing Model for Learning}
After choosing a risky path $i\in\{1,\cdots,N\}$ to travel, in practice a user may not obtain the whole path information when making his local observation and reporting to the crowdsourcing platform. 
Two different users traveling on the same path may have different experiences. Similar to \cite{li2019recommending}, we model $\alpha_i(t)$ dynamics as the partially observable two-state Markov chain in Fig.~\ref{fig:POMDP} from the user point of view. 
We define a random observation set $\mathbf{y}(t)=\{y_1(t),\cdots,y_N(t)\}$ for $N$ risky paths, where $y_i(t)\in\{0,1,\emptyset\}$ denotes the traffic condition of path $i$ as observed by the current user there during time slot~$t$. More specifically:
\begin{itemize}
    \item $y_i(t)=1$ tells that the current user arrival observes a hazard (e.g., ‘black ice’ segments, poor visibility, jamming) after choosing path $\pi(t)=i$ at time $t$.
    \item $y_i(t)=0$ tells that the current user arrival does not observe any hazard on path $\pi(t)=i$. 
    \item $y_i(t)=\emptyset$ indicates the absence of any observation for path $i$, which can happen when no user arrives with $s(t)=0$ or the user arrival travels on another path with $\pi(t)\neq i$.
\end{itemize}

Given $\pi(t)=i$ under $s(t)=1$, the chance for the user to observe $y_i(t)=1$ or $0$ depends on the random correlation coefficient $\alpha_i(t)$. Under the correlation state $\alpha_i(t)=\alpha_H^i$ or $\alpha_L^i$ at time $t$, we respectively denote the probabilities for the user to observe a hazard as:
\begin{equation}
    \begin{aligned}
        p_H&=\text{Pr}\big(y_i(t)=1|\alpha_i(t)=\alpha_H^i\big), \\
        p_L&=\text{Pr}\big(y_i(t)=1|\alpha_i(t)=\alpha_L^i\big).
    \end{aligned}\label{p_H}
\end{equation}
Note that $p_L<p_H$ because a risky path in bad traffic conditions ($\alpha_i(t)=\alpha_H^i$) has a larger probability for the user to observe a hazard (i.e., $y_i(t)=1$). Even if path $i$ has good traffic conditions ($\alpha_i(t)=\alpha_L^i$), it is not entirely hazard-free and there is still some probability $p_L$ to face a hazard.

As users keep learning and sharing traffic conditions with the crowdsourcing platform, the historical data of their observations $(\mathbf{y}(1),\cdots,\mathbf{y}(t-1))$ and routing decisions $(\pi(1),\cdots,\pi(t-1))$ before time $t$ keep growing in the time horizon. To simplify the ever-growing history set, we equivalently translate these historical observations into a hazard belief $x_i(t)$ for seeing bad traffic condition $\alpha_i(t)=\alpha_H^i$ at time $t$, by using the Bayesian inference:
\begin{equation}
    x_i(t)=\text{Pr}\big(\alpha_i(t)=\alpha_H^i|x_i(t-1),\pi(t-1),\mathbf{y}(t-1)\big).\label{def_x}
\end{equation}
Given the prior probability $x_i(t)$, the platform will further update it to a posterior probability $x_i'(t)$ after a new user with routing decision $\pi(t)$ shares his observation $y_i(t)$ during the time slot:
\begin{equation}
    x_i'(t)=\text{Pr}\big(\alpha_i(t)=\alpha_H^i|x_i(t),\pi(t),\mathbf{y}(t)\big).\label{def_x'}
\end{equation}
Below, we explain the dynamics of our information learning model.
\begin{itemize}
    \item At the beginning of time slot $t$, the platform publishes any risky path $i$’s hazard belief $x_i(t)$ in (\ref{def_x}) about coefficient $\alpha_i(t)$ and the latest expected latency $\mathbb{E}[\ell_i(t)|x_i(t-1),y_i(t-1)]$ to summarize observation history $(\mathbf{y}(1),\cdots,\mathbf{y}(t-1))$ till $t-1$.

    \item During time slot $t$, a user arrives to choose a path (e.g., $\pi(t)=i$) to travel and reports his following observation $y_i(t)$. Then the platform updates the posterior probability $x_i'(t)$, conditioned on the new observation $y_i(t)$ and the prior probability $x_i(t)$ in (\ref{def_x'}). For example, if $y_i(t)=0$, by Bayes’ Theorem, $x_i'(t)$ for the correlation coefficient $\alpha_i(t)=\alpha_H^i$ is
    \begin{align}
        x_i'(t)
        =&\text{Pr}\big(\alpha_i(t)=\alpha_H^i|x_i(t),\pi(t)=i,y_i(t)=0\big)\label{x_i'(t)_y=0}\\
        =&\frac{x_i(t)(1-p_H)}{x_i(t)(1-p_H)+(1-x_i(t))(1-p_L)}.\notag
    \end{align}
    Similarly, if $y(t)=1$, we have
    \begin{equation}
        x_i'(t)=\frac{x_i(t)p_H}{x_i(t)p_H+(1-x_i(t))p_L}.\label{x_i'(t)_y=1}
    \end{equation}
    Besides this traveled path $i$, for any other path $j\in\{1,\cdots,N\}$ with $y_j(t)=\emptyset$, we keep $x'_j(t)=x_j(t)$ as there is no added observation to this path at $t$.

    \item At the end of this time slot, the platform estimates the posterior correlation coefficient:
    \begin{equation}
        \begin{aligned}
            \mathbb{E}[\alpha_i(t)|x'_i(t)]&=\mathbb{E}[\alpha_i(t)|x_i(t),y_i(t)]
            \\&=x'_i(t)\alpha_H^i+(1-x'_i(t))\alpha_L^i.
        \end{aligned}\label{E_alpha}
    \end{equation} 
    By combining (\ref{E_alpha}) with (\ref{L_i(t+1)}), we can obtain the expected travel latency on stochastic path $i$ for time $t+1$ as
    \begin{align}
        &\Scale[0.95]{\mathbb{E}[\ell_i(t+1)|x_i(t),y_i(t)]}\label{E[L_i(t+1)]}=\\
        \!\!&\begin{cases}
            \Scale[0.95]{\mathbb{E}[\alpha_i(t)|x'_i(t)] \mathbb{E}[\ell_i(t)|x_i(t-1),y_i(t-1)]+\Delta \ell},\\
            \quad\quad\quad\quad\quad\quad\quad\quad\quad\quad\quad\quad\quad\quad\quad\quad\ \text{ if }\Scale[0.95]{\pi(t)=i},\\
            \Scale[0.95]{\mathbb{E}[\alpha_i(t)|x'_i(t)] \mathbb{E}[\ell_i(t)|x_i(t-1),y_i(t-1)]},\text{ if }\Scale[0.95]{\pi(t)\neq i}.
        \end{cases}\notag
    \end{align}
    Based on the partially observable Markov chain in Fig.~\ref{fig:POMDP}, the platform updates each path $i$'s hazard belief from $x_i'(t)$ to $x_i(t+1)$ below: 
    \begin{equation}
        x_i(t+1)=x_i'(t)q_{HH}^i+\big(1-x_i'(t)\big)q_{LH}^i.\label{x_i(t)}
    \end{equation}
    Finally, the new time slot $t+1$ begins and repeats the process as described above.
\end{itemize}

\section{POMDP Problem Formulations for Myopic and Socially Optimal Policies}\label{section3}
Based on the dynamic congestion and crowdsourcing models in the last section, we formulate the problems of myopic policy (for guiding myopic users' selfish routing) and the socially optimal policy (for the social planner/platform's best path advisory), respectively. 

\subsection{Problem Formulation for Myopic Policy}
In this subsection, we consider the myopic policy (e.g. used by Waze and Google Maps) that selfish users will naturally follow. 
First, we summarize the dynamics of expected travel latencies among all $N+1$ paths and the hazard beliefs of $N$ stochastic paths into vectors:
\begin{align}
    &\begin{aligned}
        \mathbf{L}(t)=\big\{&\ell_0(t),\mathbb{E}[\ell_1(t)|x_1(t-1),y_1(t-1)],\cdots,\\&\mathbb{E}[\ell_N(t)|x_N(t-1),y_N(t-1)]\big\},
    \end{aligned}\notag\\
    &\mathbf{x}(t)=\{x_1(t),\cdots,x_N(t)\},\label{LX_set}
\end{align}
which are obtained based on (\ref{E[L_i(t+1)]}) and (\ref{x_i(t)}). For a user arrival at time $t$, the platform provides him with $\mathbf{L}(t)$ and $\mathbf{x}(t)$ to help make his routing decision. 
We define the best stochastic path~$\hat{\iota}(t)$ to be the one out of $N$ risky paths to provide the shortest expected travel latency at time $t$ below:
\begin{equation}
    \hat{\iota}(t)=\arg\min_{i\in\{1,\cdots,N\}} \mathbb{E}[\ell_i(t)|x_i(t-1),y_i(t-1)]. \label{hat_i}
\end{equation}
The selfish user will only choose between safe path 0 and this path $\hat{\iota}(t)$ to minimize his own travel latency. 

We formulate this problem as a POMDP, where the time correlation state $\alpha_i(t)$ of each stochastic path $i$ is partially observable to users in Fig. \ref{fig:POMDP}. 
Thus, the states here are $\mathbf{L}(t)$ and $\mathbf{x}(t)$ in (\ref{LX_set}). Under the myopic policy, define ${C^{(m)}\big(\mathbf{L}(t), \mathbf{x}(t),s(t)\big)}$ to be the long-term discounted cost function with discount factor $\rho<1$ to include the social cost of all users since $t$. 

If $s(t)=1$ with user arrival, then its dynamics per user arrival has the following two cases.
If $\mathbb{E}[\ell_{\hat{\iota}(t)}(t)|x_{\hat{\iota}(t)}(t-1),y_{\hat{\iota}(t)}(t-1)]\geq \ell_0(t)$, a selfish user will choose path 0 and add $\Delta \ell$ to path 0 to have latency $\ell_0(t+1)=\alpha \ell_0(t)+\Delta \ell$ in (\ref{L_0(t+1)}). Since no user enters stochastic path $i$, there is no information reporting (i.e., $y_i(t)=\emptyset$) and $x_i'(t)$ in (\ref{def_x'}) equals $x_i(t)$ in (\ref{def_x}) for updating $x_i(t+1)$ in (\ref{x_i(t)}). 
The expected travel latency of stochastic path $i$ in the next time slot is updated to $\mathbb{E}[\ell_i(t+1)|x_i(t),y_i(t)=\emptyset]$ according to (\ref{E[L_i(t+1)]}). In consequence, the travel latency and hazard belief sets at the next time slot $t+1$ are updated to
\begin{align}
    &\begin{aligned}\mathbf{L}(t+1)=\big\{&\alpha\ell_0(t)+\Delta\ell, \mathbb{E}[\ell_1(t+1)|x_1(t),y_1(t)=\emptyset],\\&\cdots,\mathbb{E}[\ell_N(t+1)|x_N(t),y_N(t)=\emptyset]\big\},\end{aligned}\notag\\
    &\mathbf{x}(t+1)=\big\{x_1(t+1),\cdots,x_N(t+1)\big\}.\label{L0X0}
\end{align}
Then the cost-to-go $Q_0^{(m)}(t+1)$ since the next time slot is
\begin{equation}
    \Scale[0.95]{Q_0^{(m)}(t+1)=C^{(m)}\Big(\mathbf{L}(t+1),\mathbf{x}(t+1),s(t+1)\big|y_{\hat{\iota}(t)}(t)=\emptyset\Big)}.\label{Q_0m}
\end{equation}

If $\Scale[0.95]{\mathbb{E}[\ell_{\hat{\iota}(t)}(t)|x_{\hat{\iota}(t)}(t-1),y_{\hat{\iota}(t)}(t-1)]< \ell_0(t)}$, the user arrival will choose the best stochastic path $\hat{\iota}(t)$ in (\ref{hat_i}). Then the platform updates the expected travel latency on path $\hat{\iota}(t)$ to $\Scale[0.95]{\mathbb{E}[\ell_{\hat{\iota}(t)}(t)|x_{\hat{\iota}(t)}(t),y_{\hat{\iota}(t)}(t)]}$ in (\ref{E[L_i(t+1)]}), depending on whether $y_{\hat{\iota}(t)}(t)=1$ or $0$. Note that according to (\ref{p_H}),
\begin{equation}
    \Scale[0.95]{\mathbf{Pr}\big(y_{\hat{\iota}(t)}(t)=1\big)=\big(1-x_{\hat{\iota}(t)}(t)\big)p_L+x_{\hat{\iota}(t)}(t)p_H}. \label{PG}
\end{equation}
While path 0's latency in next time changes to $\alpha\ell_0(t)$, and path $i\neq\hat{\iota}(t)$ has no exploration and its expected latency at time $t+1$ becomes $\mathbb{E}[\ell_i(t+1)|x_i(t),y_i(t)=\emptyset]$. Then the expected cost-to-go since the next time slot in this case is
\begin{align}
    &\Scale[0.92]{Q_{\hat{\iota}(t)}^{(m)}(t+1)=} \label{E[C]} \\ &\Scale[0.92]{\mathbf{Pr}\big(y_{\hat{\iota}(t)}(t)=1\big)C^{(m)}\Big(\mathbf{L}(t+1),\mathbf{x}(t+1),s(t+1)\big|y_{\hat{\iota}(t)}(t)=1\Big)+}\notag\\&\Scale[0.92]{\mathbf{Pr}\big(y_{\hat{\iota}(t)}(t)=0\big)C^{(m)}\Big(\mathbf{L}(t+1),\mathbf{x}(t+1),s(t+1)\big|y_{\hat{\iota}(t)}(t)=0\Big)}.\notag
\end{align}

To combine (\ref{Q_0m}) and (\ref{E[C]}), we formulate the $\rho$-discounted long-term cost function with a user arrival at time $t$ (i.e., $s(t)=1$) under the myopic policy as
\begin{align}
    &\Scale[0.95]{C^{(m)}\big(\mathbf{L}(t), \mathbf{x}(t),1\big)=}\label{cost_Cm_s1}\\
    &\begin{cases}\notag
     \Scale[0.97]{\ell_0(t)+\rho Q_{0}^{(m)}(t+1)},\\ \quad\quad\quad\quad\quad\ \ \text{if }\Scale[0.97]{\mathbb{E}[\ell_{\hat{\iota}(t)}(t)|x_{\hat{\iota}(t)}(t-1),y_{\hat{\iota}(t)}(t-1)]\geq \ell_0(t)},\\
     \Scale[0.97]{\mathbb{E}[\ell_{\hat{\iota}(t)}(t)|x_{\hat{\iota}(t)}(t-1),y_{\hat{\iota}(t)}(t-1)]+\rho Q_{\hat{\iota}(t)}^{(m)}(t+1)},\\\quad\quad\quad\quad\quad\quad\quad\quad\quad\quad\quad\quad\quad\quad\quad\quad\quad\quad\quad\ \ \text{otherwise.}
    \end{cases}
\end{align}
According to (\ref{cost_Cm_s1}), a new arriving selfish user is not willing to explore any stochastic path $i$ with longer expected travel latency when he arrives, and the next arrival may not know the fresh congestion information. On the other hand, selfish users may keep choosing the path with the shortest latency and jamming this path for future users. 

If $s(t)=0$ without user arrival, there is no observation for any path, and we obtain the resulting long-term cost function as follows
\begin{align}\label{cost_Cm_s0}
    C^{(m)}\big(\mathbf{L}(t),\mathbf{x}(t),0\big)=\rho Q_{\emptyset}^{(m)}(t+1),
\end{align}
where the cost-to-go $Q_{\emptyset}^{(m)}(t+1)$ is similarly defined as in~(\ref{Q_0m}). 
Based on the two possible long-term cost functions above, we finally formulate the general $\rho$-discounted long-term cost function since time $t$ under the myopic policy as
\begin{align}\label{cost_Cm}
    C^{(m)}\big(\mathbf{L}(t),\mathbf{x}(t),s(t)\big)=\begin{cases}
        C^{(m)}\big(\mathbf{L}(t),\mathbf{x}(t),0\big), \text{ if }s(t)=0,\\
        C^{(m)}\big(\mathbf{L}(t),\mathbf{x}(t),1\big),\text{ if }s(t)=1,
    \end{cases}
\end{align}
where $C^{(m)}\big(\mathbf{L}(t),\mathbf{x}(t),0\big)$ and $C^{(m)}\big(\mathbf{L}(t),\mathbf{x}(t),1\big)$ are defined in (\ref{cost_Cm_s0}) and (\ref{cost_Cm_s1}), respectively. By observing (\ref{cost_Cm}), we find that the expected social cost is smaller than (\ref{cost_Cm_s1}), due to the possibility of no user arrival with $s(t)=0$ at current time~$t$. 

\subsection{Socially Optimal Policy Problem Formulation}
Different from the myopic policy that focuses on the one-shot to minimize the current user's immediate travel cost if he arrives, the goal of the social optimum is to find optimal policy $\pi^*(t)$ at any time $t$ to minimize the expected social cost over an infinite time horizon.

Denote the long-term $\rho$-discounted cost function by ${C^*\big(\mathbf{L}(t), \mathbf{x}(t),s(t)\big)}$ under the socially optimal policy. The optimal policy depends on which path choice yields the minimal long-term social cost. We first analyze the case with user arrival $s(t)=1$ at time $t$. If the platform asks the current user to choose path 0, this user will bear the cost $\ell_0(t)$ to travel this path. Due to no information observation (i.e., $\mathbf{y}(t)=\emptyset$), the cost-to-go $Q^*_0(t+1)$ from the next time slot can be similarly determined as (\ref{Q_0m}) with $\mathbf{L}(t+1)$ and $\mathbf{x}(t+1)$ in (\ref{L0X0}).

If the platform asks the user to explore a stochastic path~$i$, this choice is not necessarily path $\hat{\iota}(t)$ in (\ref{hat_i}). Then the platform updates $\mathbf{x}(t+1)$, depending on whether the user's observation on this path is $y_i(t)=1$ or $y_i(t)=0$. Similar to (\ref{E[C]}), the optimal expected cost function from the next time slot is denoted as $Q^*_{i}(t+1)$. Then we are ready to formulate the long-term cost function with $s(t)=1$ under the socially optimal policy below:
\begin{align}
    &C^*\big(\mathbf{L}(t), \mathbf{x}(t),1\big)\label{cost_C*_s1}\\=&\min_{i\in\{1,\cdots,N\}}\big\{\ell_0(t)+\rho Q^*_0(t+1), \ell_{i}(t)+\rho Q^*_{i}(t+1)\big\}.\notag
\end{align}

If there is no user arrival at time $t$, i.e., $s(t)=0$, the resulting long-term social cost $C^*(\mathbf{L}(t), \mathbf{x}(t),0)$ is similarly defined as in (\ref{cost_Cm_s0}). Then we obtain the long-term cost function in the general case under the socially optimal policy as
\begin{align}\label{cost_C*}
    C^*\big(\mathbf{L}(t),\mathbf{x}(t),s(t)\big)=\begin{cases}
        C^*\big(\mathbf{L}(t),\mathbf{x}(t),0\big), \text{ if }s(t)=0,\\
        C^*\big(\mathbf{L}(t),\mathbf{x}(t),1\big),\text{ if }s(t)=1.
    \end{cases}
\end{align}
Problem (\ref{cost_C*}) is non-convex and its analysis will cause the curse of dimensionality in the infinite time horizon \cite{bellman1966dynamic}.
Though it is difficult to solve, we manage to analytically compare the two policies by investigating their structural results below.

\section{Comparing Myopic Policy to Social Optimum for PoA Analysis}\label{section4}
In this section, we first prove that both myopic and socially optimal policies to explore stochastic paths are of threshold-type with respect to expected travel latency. Then we show that the myopic policy may both under-explore and over-explore risky paths.\footnote{Over/under exploration means that myopic policy will choose risky path~$i$ more/less often than what the social optimum suggests.} Finally, we prove that the myopic policy can perform arbitrarily bad.

\begin{lemma}\label{lemma:monotonocity}
The cost functions $C^{(m)}\big(\mathbf{L}(t), \mathbf{x}(t),s(t)\big)$ in (\ref{cost_Cm}) and $C^*\big(\mathbf{L}(t), \mathbf{x}(t),s(t)\big)$ in (\ref{cost_C*}) under both policies increase with any path's expected latency $\mathbb{E}[\ell_i(t)|x_i(t-1),y_i(t-1)]$ in $\mathbf{L}(t)$ and $\mathbf{x}(t)$ in (\ref{LX_set}).
\end{lemma}
The proof of Lemma \ref{lemma:monotonocity} is given in Appendix A. With this monotonicity result, we next prove that both policies are of threshold-type for the possible user arrival at time $t$.
\begin{proposition}\label{prop:threshold}
Provided with $\mathbf{L}(t)$ and $\mathbf{x}(t)$ in (\ref{LX_set}), the possible user arrival at time $t$ under the myopic policy keeps staying with path 0, until the expected latency of the best stochastic path $\hat{\iota}(t)$ in (\ref{hat_i}) reduces to be smaller than the following threshold:
\begin{equation}
    \ell^{(m)}(t)=\ell_0(t). \label{L_mopic}
\end{equation}
Similarly, the socially optimal policy will choose stochastic path $i$ instead of path 0 if $\mathbb{E}[\ell_i(t)|x_i(t-1),y_i(t-1)]$ is less than the following threshold:
\begin{equation}
    \begin{aligned}
    \Scale[0.95]{\ell_i^*(t)=\arg \max_{z}\big\{z|z \leq \rho Q^*_{i}(t+1)-\rho Q^*_0(t+1)-\ell_0(t)\big\},}
    \end{aligned}\label{L_optimal}
\end{equation}
which increases with hazard belief $x_{i}(t)$ of risky path $i$.
\end{proposition}

The proof of Proposition \ref{prop:threshold} is given in Appendix B. Let $\pi^{(m)}(t)$ and $\pi^*(t)$ denote the routing decisions at time $t$ under myopic and socially optimal policies, respectively. We next compare the exploration thresholds $\ell^{(m)}(t)$ and $\ell_i^{*}(t)$ as well as their associated social costs. 
\begin{lemma}\label{lemma:explore}
If $\pi^{(m)}(t)\neq \pi^*(t)$, then the expected travel latencies on these two chosen paths by the two policies satisfy
\begin{align}
    &\mathbb{E}[\ell_{\pi^*(t)}(t)|\mathbf{x}(t-1),\mathbf{y}(t-1)]\notag\\ \leq& \frac{1}{1-\rho}\mathbb{E}[\ell_{\pi^{(m)}(t)}(t)|\mathbf{x}(t-1),\mathbf{y}(t-1)].\label{lemma_inequality}
\end{align}
\end{lemma}

The proof of Lemma \ref{lemma:explore} is given in Appendix C. Intuitively, if the current travel latencies on different paths obviously differ, the two policies tend to make the same routing decision. (\ref{lemma_inequality}) is more likely to hold for large $\rho$.

Next, we define the stationary belief of high hazard state $\alpha_H^i$ as $\Bar{x}^i$, and we provide it below by using steady-state analysis of the Markov chain in Fig. \ref{fig:POMDP}:
\begin{equation}
    \Bar{x}_i=\frac{1-q_{LL}^i}{2-q_{LL}^i-q_{HH}^i}.\label{x_bar}
\end{equation}
Based on Proposition \ref{prop:threshold} and Lemma \ref{lemma:explore}, we analytically compare the two policies below.
\begin{proposition}\label{prop:explore}
There exists a belief threshold $x^{th}$ satisfying
\begin{equation}
    \min\Big\{\frac{\alpha-\alpha^i_L}{\alpha^i_H-\alpha^i_L},\Bar{x}_i\Big\}\leq x^{th}_i\leq \max\Big\{\frac{\alpha-\alpha^i_L}{\alpha^i_H-\alpha^i_L},\Bar{x}_i\Big\}.\label{x_th}
\end{equation}
As compared to socially optimal policy, if risky path $i\in\{1,\cdots,N\}$ has weak hazard belief $x_i(t)< x^{th}_i$, myopic users will only over-explore this path with $\ell^{(m)}(t)\geq \ell_i^*(t)$. If strong hazard belief with $x_i(t)>x^{th}_i$, myopic users will only under-explore this path with $\ell^{(m)}(t)\leq \ell_i^*(t)$.
\end{proposition}

The proof of Proposition \ref{prop:explore} is given in Appendix D. Here $\frac{\alpha-\alpha_L^i}{\alpha_H^i-\alpha_L^i}$ in (\ref{x_th}) is derived by equating path $i$'s expected coefficient $\mathbb{E}[\alpha_i(t)|x'_i(t)]$ in (\ref{E_alpha}) to path 0's $\alpha$. Proposition \ref{prop:explore} tells that the myopic policy misses both exploitation and exploration over time. If the hazard belief on path $i\in\{1,\cdots,N\}$ is weak (i.e., $x_i(t)<x^{th}_i$), myopic users choose stochastic path $i$ without considering the congestion to future others on the same path. While the socially optimal policy may still recommend users to safe path 0 to further reduce the congestion cost on path $i$ for the following user. On the other hand, if $x_i(t)>x^{th}_i$, the socially optimal policy may still want to explore path $i$ to exploit hazard-free state $\alpha_L$ on this path for future use. This result is also consistent with $\ell_i^*(t)$'s monotonicity in $x_i(t)$ in Proposition \ref{prop:threshold}. 

\begin{figure}[t]
    \centering
    \includegraphics[width=0.35\textwidth]{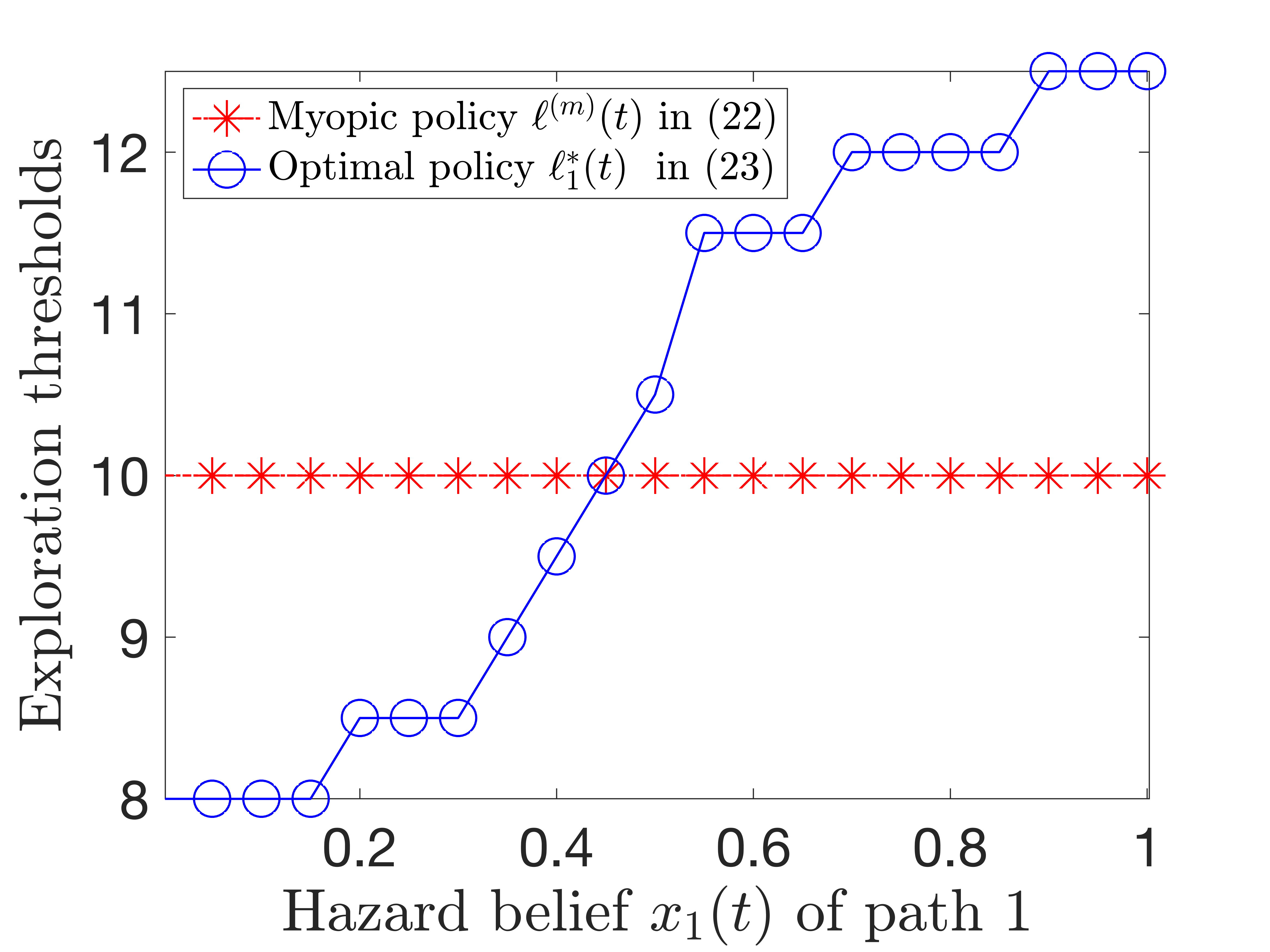}
    \caption{The socially optimal policy's exploration threshold $\ell_1^*(t)$ and myopic policy's threshold $\ell^{(m)}(t)$ versus hazard belief $x_1(t)$ in a two-path transportation network with $N=1$.  We set $\alpha=0.6,\alpha_H^1=1.2,\alpha_L^1=0.2,q_{LL}^1=0.5,q_{HH}^1=0.5,\Delta\ell=2,p_H=0.8,p_L=0.3$ and $\ell_0(t)=10$ at current time $t$.}
    \label{fig:exploration_x}
\end{figure}

In Fig. \ref{fig:exploration_x}, we simulate Fig. \ref{fig:congestion_game} using a simple two-path transportation network with $N=1$. We plot exploration thresholds $\ell^{(m)}(t)$ in (\ref{L_mopic}) under the myopic policy and optimal $\ell_1^*(t)$ in (\ref{L_optimal}) versus hazard belief $x_1(t)$ of path 1. These two thresholds are very different in Fig. \ref{fig:exploration_x}. 
Given the belief threshold $x^{th}_1 = 0.45$ (where $\ell^{(m)}(t)$ and $\ell_1^(t)$ are equal at $x_1(t) = 0.45$), if the hazard belief $x_1(t) < x^{th}_1$, the myopic exploration threshold $\ell^{(m)}(t)$ exceeds $\ell_1^(t)$, leading to over-exploration of the stochastic path. 
If $x_1(t)>x^{th}_1$, the myopic exploration threshold satisfies $\ell^{(m)}(t)<\ell_1^*(t)$ to under-explore. This result is consistent with Proposition \ref{prop:explore}.

After comparing the two policies' thresholds, we are ready to further examine their performance gap. Following \cite{koutsoupias1999worst}, we define the price of anarchy (PoA) to be the maximum ratio between the social cost under the myopic policy in (\ref{cost_Cm}) and the minimal social cost in (\ref{cost_C*}), by searching all possible system parameters:
\begin{align}
    &\text{PoA}^{(m)}=\max_{\substack{\lambda,\alpha,\alpha_H^i,\alpha_L^i,q_{LL}^i,q_{HH}^i,\\\mathbf{x}(t),\mathbf{L}(t),\Delta \ell, p_H, p_L}}{\frac{C^{(m)}\big(\mathbf{L}(t), \mathbf{x}(t),s(t)\big)}{C^{*}\big(\mathbf{L}(t), \mathbf{x}(t,s(t))\big)}},\label{PoA}
\end{align}
which is obviously larger than 1. Then we present the lower bound of PoA in the following proposition.
\begin{proposition}\label{thm:poa}
As compared to the social optimum in (\ref{cost_C*}), the myopic policy in (\ref{cost_Cm}) achieves $\text{PoA}^{(m)}\geq \frac{1}{1-\rho^{\frac{1}{\lambda}}}$, which increases with $\lambda$ and can be arbitrarily large for discount factor $\rho\rightarrow 1$.
\end{proposition}

The proof of Proposition \ref{thm:poa} is given in Appendix E. In this worst-case PoA analysis, we consider a two-path network example, where the myopic policy always chooses safe path 0 but the socially optimal policy frequently explores stochastic path 1 to learn $\alpha_L^i$. 
Here we initially set $\ell_0(0)=\frac{\Delta \ell}{1-\alpha^{\frac{1}{\lambda}}}$ such that the expected travel latency $\ell_0(\frac{1}{\lambda})=\alpha^{\frac{1}{\lambda}}\ell_0(0)+\Delta \ell$ in (\ref{L_0(t+1)}) equals $\frac{\Delta\ell}{1-\alpha^{\frac{1}{\lambda}}}$ all the time for myopic users. Without myopic users' routing on stochastic path 1, we also keep the expected travel latency on stochastic path 1 unchanged, by setting $x_1(0)=\Bar{x}$ in (\ref{x_bar}) and $\mathbb{E}[\alpha_1(0)|x_1(0)=\Bar{x}]=1$ in (\ref{E_alpha}). Then a myopic user at any time $t$ will never explore the stochastic path 1 given $\ell_1(t)=\ell_0(t)$, resulting in the social cost to be $\frac{\ell_0(0)}{1-\rho^{\frac{1}{\lambda}}}$ with expected arrival interval $\frac{1}{\lambda}$ in the infinite time horizon.
However, the socially optimal policy frequently asks a user to explore path 1 to learn a good condition ($\alpha_L^i=0$) for following users when he arrives. We make $q_{LL}^i\rightarrow 1$ to maximally reduce the travel latency of path 1, and the optimal social cost is thus no more than $\rho^{\frac{1}{\lambda}}\ell_1(0)+\frac{\rho^{\frac{2}{\lambda}}}{1-\rho^{\frac{1}{\lambda}}}\Delta\ell$. Letting $\frac{\Delta\ell}{\ell_0(0)}\rightarrow 0$, we obtain $\text{PoA}^{(m)}\geq \frac{1}{1-\rho^{\frac{1}{\lambda}}}$. 

This $\text{PoA}^{(m)}$ increases with arrival probability~$\lambda$ because a higher arrival probability leads to more selfish user arrivals, which, in turn, increases the total cost under the myopic policy. By Proposition \ref{thm:poa}, the myopic policy performs worse, as discount factor $\rho$ increases and future costs become more important. As $\rho\rightarrow 1$ and $\lambda>0$, PoA approaches infinity and the learning efficiency in the crowdsourcing platform
becomes arbitrarily bad to opportunistically reduce the congestion. Thus, it is critical to design an efficient incentive mechanism to greatly reduce the social cost.

\section{Selective Information Disclosure}\label{section5}
To motivate a selfish user to follow the optimal path advisory when he arrives, we need to design a non-monetary information mechanism, which naturally satisfies budget balance and is easy to implement without enforcing monetary payments. Our key idea is to selectively disclose the latest expected travel latency set $\mathbf{L}(t)$ of all paths, depending on a myopic user's intention to over- or under-explore stochastic paths at time $t$. To avoid users from perfectly inferring $\mathbf{L}(t)$, we purposely hide the latest hazard belief set $\mathbf{x}(t)$, routing history $\big(\pi(1),\cdots,\pi(t-1)\big)$, and past traffic observation set $\big(\mathbf{y}(1),\cdots, \mathbf{y}(t-1)\big)$, but always provide socially optimal path recommendation $\pi^*(t)$ to any user.
Provided with selective information disclosure, we allow sophisticated users to reverse-engineer the path latency distribution and make selfish routing under our mechanism.
For simplicity, we assume $\alpha_H^i=\alpha_H$, $\alpha_L^i=\alpha_L$, $q_{HH}^i=q_{HH}$ and $q_{LL}^i=q_{LL}$ for any risky path $i$ in this section. 
However, our SID mechanism is also applicable to the general case, as verified by the real-data experiments presented later in Section~\ref{section7}.

Before formally introducing our selective information disclosure in Definition \ref{def:mechanism}, we first consider an information hiding policy $\pi^{\emptyset}(t)$ as a benchmark. Similar information-hiding mechanisms were proposed and studied in the literature (e.g., \cite{tavafoghi2017informational} and \cite{li2019recommending}). In this benchmark mechanism, the user without any information believes that the expected hazard belief $x_i(t)$ of any stochastic path $i\in \{1,\cdots,N\}$ has converged to its stationary hazard belief $\Bar{x}$ in (\ref{x_bar}). Then he can only decide his routing policy $\pi^{\emptyset}(t)$ by comparing $\alpha$ of safe path 0 to $\mathbb{E}[\alpha_i(t)|\Bar{x}]$ in (\ref{E_alpha}) of any path $i$. 
\begin{proposition}\label{lemma:pi^empty}
Given no information from the platform, a user arrival at time $t$ uses the following routing policy: 
\begin{equation}
    \pi^{\emptyset}(t)=\begin{cases}
    0, &\text{if }\Bar{x}\geq \frac{\alpha-\alpha_L}{\alpha_H-\alpha_L},\\
    i\text{ w/ probability }\frac{1}{N},&\text{if }\Bar{x}< \frac{\alpha-\alpha_L}{\alpha_H-\alpha_L},
    \end{cases}\label{pi^empty}
\end{equation}
where $i\in\{1,\cdots,N\}$. This hiding policy leads to $\text{PoA}^{\emptyset}\rightarrow\infty$, regardless of discount factor $\rho$.
\end{proposition}

The proof of Proposition \ref{lemma:pi^empty} is given in Appendix F. Even if we still recommend optimal routing $\pi^*(t)$ in (\ref{cost_C*}), a selfish user sticks to some risky path $i$ given low hazard belief $\Bar{x}<\frac{\alpha-\alpha_L}{\alpha_H-\alpha_L}$. 
This hiding policy can differ a lot from the socially optimal policy in (\ref{cost_C*}) since users cannot observe the latest travel latencies. To tell the $\text{PoA}^{\emptyset}=\infty$, we consider the simplest two-path network example: initially safe path 0 has $\ell_0(t=0)=0$ with $\alpha\rightarrow 1$, and risky path 1 has an arbitrarily large travel latency $\ell_1(0)$ with $\Bar{x}=0$ and $\mathbb{E}[\alpha_1(t)|\Bar{x}]= 0$, by letting $q_{LL}=1$ and $\alpha_L=0$. 
Given $\mathbb{E}[\alpha_1(t)|\Bar{x}]<\alpha$ or simply $\Bar{x}<\frac{\alpha-\alpha_L}{\alpha_H-\alpha_L}$, a selfish user always chooses path $\pi^{\emptyset}(t)=1$ when he arrives, leading to social cost $\rho^{\frac{1}{\lambda}}\ell_1(0)+\frac{\rho^{\frac{2}{\lambda}}\Delta\ell}{1-\rho^{\frac{1}{\lambda}}}$. 
While letting the first arriving user exploit $\ell_0({\frac{1}{\lambda}})=0$ of path 0 to reduce $\mathbb{E}[\ell_1({\frac{2}{\lambda}})|\Bar{x},\emptyset]$ to 0 for path 1 at his expected arrival time ${\frac{1}{\lambda}}$, the socially optimal cost is thus $\frac{\rho^{\frac{2}{\lambda}}\Delta\ell}{1-\rho^{\frac{1}{\lambda}}}$. Letting $\frac{(1-\rho^{\frac{1}{\lambda}})\ell_1(0)}{\rho^{\frac{2}{\lambda}}\Delta\ell}\rightarrow\infty$, we obtain $\text{PoA}^{\emptyset}=\infty$. 

This is a $\text{PoA}^{\emptyset}$ example with the maximum-exploration of stochastic paths, which is opposite to the zero-exploration $\text{PoA}^{(m)}$ example after Proposition \ref{thm:poa}. 
Given neither information hiding policy $\pi^{\emptyset}(t)$ nor myopic policy $\pi^{(m)}(t)$ under full information sharing works well, we need to design an efficient mechanism to selectively disclose information to users to reduce the social cost. 
\begin{definition}[Selective Information Disclosure (SID) Mechanism]\label{def:mechanism}
If a user arrival at time $t$ is expected to choose a different route $\pi^{\emptyset}(t)\neq 0$ in (\ref{pi^empty}) from optimal $\pi^*(t)=0$ in (\ref{cost_C*}), then our SID mechanism will disclose the latest expected travel latency set $\mathbf{L}(t)$ to him. Otherwise, our mechanism hides $\mathbf{L}(t)$ from this user. Besides, our mechanism always provides optimal path recommendation $\pi^*(t)$, without sharing hazard belief set $\mathbf{x}(t)$, routing history $\big(\pi(1),\cdots,\pi(t-1)\big)$, or past observation set $\big(\mathbf{y}(1),\cdots, \mathbf{y}(t-1)\big)$.
\end{definition}

According to Definition \ref{def:mechanism}, if $\pi^*(t)=0$ but a user at time $t$ makes routing decision $\pi^{\emptyset}(t)\neq 0$ under $\Bar{x}<\frac{\alpha-\alpha_L}{\alpha_H-\alpha_L}$ in (\ref{pi^empty}), our mechanism discloses $\mathbf{L}(t)$ to avoid him from choosing any stochastic path with large expected travel latency. In the other cases, we simply hide $\mathbf{L}(t)$ from any user arrival, as the user already follows optimal routing $\pi^*(t)$. 

In consequence, the worst-case for our SID mechanism only happens when $\pi^{\emptyset}(t)\neq 0$ and $\pi^*(t)=0$ under $\Bar{x}<\frac{\alpha-\alpha_L}{\alpha_H-\alpha_L}$ in (\ref{pi^empty}). 
We still consider the same two-path network example with the maximum exploration after Proposition \ref{lemma:pi^empty} to show why this SID mechanism works. In this example, our mechanism will provide $\mathbf{L}(t)$, including $\ell_0(0)$ and $\ell_1(0)$, to each user arrival. Observing huge $\ell_1(0)$, the first user turns to choose path $0$ with $\ell_0(0)=0$, which successfully avoids the infinite social cost under $\pi^{\emptyset}(t)$. 
Furthermore, our SID mechanism successfully avoids the worst case of $\text{PoA}^{(m)}$ in Proposition~\ref{thm:poa}. Next, we prove that our mechanism well bounds the PoA in the following.
\begin{theorem}\label{thm:poa_incentive}
Our SID mechanism results in $\text{PoA}^{(\text{SID})}\leq \frac{1}{1-\frac{\rho^{\frac{1}{\lambda}}}{2}}$, which is always no more than $2$.
\end{theorem}

The proof of Theorem \ref{thm:poa_incentive} is given in Appendix G. In the worst-case of $\pi^{\emptyset}(t)\neq 0$ and $\pi^*(t)=0$ for our SID mechanism's $\text{PoA}^{(\text{SID})}$, a user knowing $\mathbf{L}(t)$ may deviate to follow the myopic policy $\pi^{(m)}(t)\neq 0$ in (\ref{cost_Cm}).
To explain the bounded $\text{PoA}^{(\text{SID})}$, we consider a two-path network example with the maximum exploration under the myopic policy. 
Here we start with $\ell_0(0)=\ell_1(0)-\varepsilon$ for safe path 0 with $\alpha\rightarrow 1$ to keep the travel latency on path 0 unchanged if no user chooses that path, where $\varepsilon$ is positive infinitesimal. 
We set $\ell_1(0)=\frac{\Delta \ell}{1-\mathbb{E}[\alpha_1(0)|\Bar{x}]^{\frac{1}{\lambda}}}$ for stochastic path 1 with $x_1(0)=\Bar{x}$, such that the travel latency $\mathbb{E}[\ell_1(t)|\Bar{x},y_1(t-1)]$ equals $\ell_1(0)$ all the time if all users choose that path.
Then in this system, users keep choosing path~1 under myopic policy $\pi^{(m)}(t)$ in (\ref{cost_Cm}) to receive social cost $\frac{\rho^{\frac{1}{\lambda}}\ell_1(0)}{1-\rho^{\frac{1}{\lambda}}}$. 
However, the socially optimal policy may want the first user to exploit path 0 to permanently reduce path 1's expected travel latency for following users there. Thanks to the first user's routing of path 0, the expected travel latency for each following user choosing path 1 at time $t$ is greatly reduced to be less than $\ell_1(0)$ yet is still no less than $\frac{\ell_1(0)}{2}$ for non-zero $\mathbb{E}[\alpha_1(t)|\Bar{x}]$. Then the minimum social cost is reduced to be no less than $\rho^{\frac{1}{\lambda}}\ell_0(0)+\frac{\rho^{\frac{2}{\lambda}} \ell_1(0)}{2-2\rho^{\frac{1}{\lambda}}}$, leading to $\text{PoA}^{\text{(SID)}}\leq \frac{1}{1-\frac{\rho^{\frac{1}{\lambda}}}{2}}$.

\begin{figure}[t]
    \centering
    \captionsetup{font={footnotesize}}
    \subfigure[We vary risky path number $N$ in set $\{2,3,4,5\}$, and we change $\alpha_H=2$ and $\alpha_H=5$ to make a comparison.]{\label{fig:avg_ratio}\includegraphics[width=0.34\textwidth]{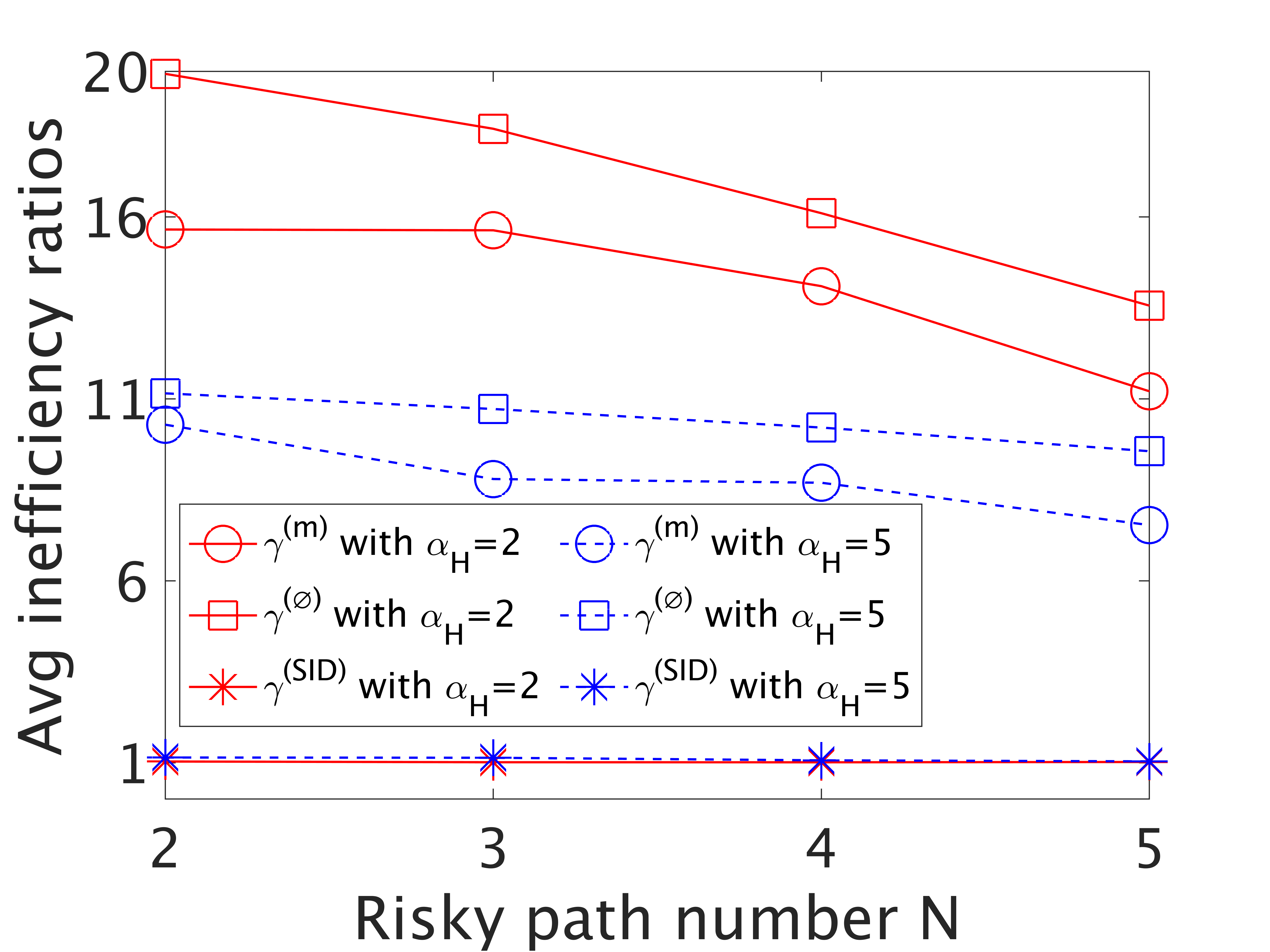}} 
    \subfigure[We vary users' arrival probability $\lambda$ in set $\{0.2,0.4,0.6,0.8,1.0\}$, and we change $\rho=0.99$ and $\rho=0.6$ to make a comparison.]{\label{fig:rate_comparison}\includegraphics[width=0.34\textwidth]{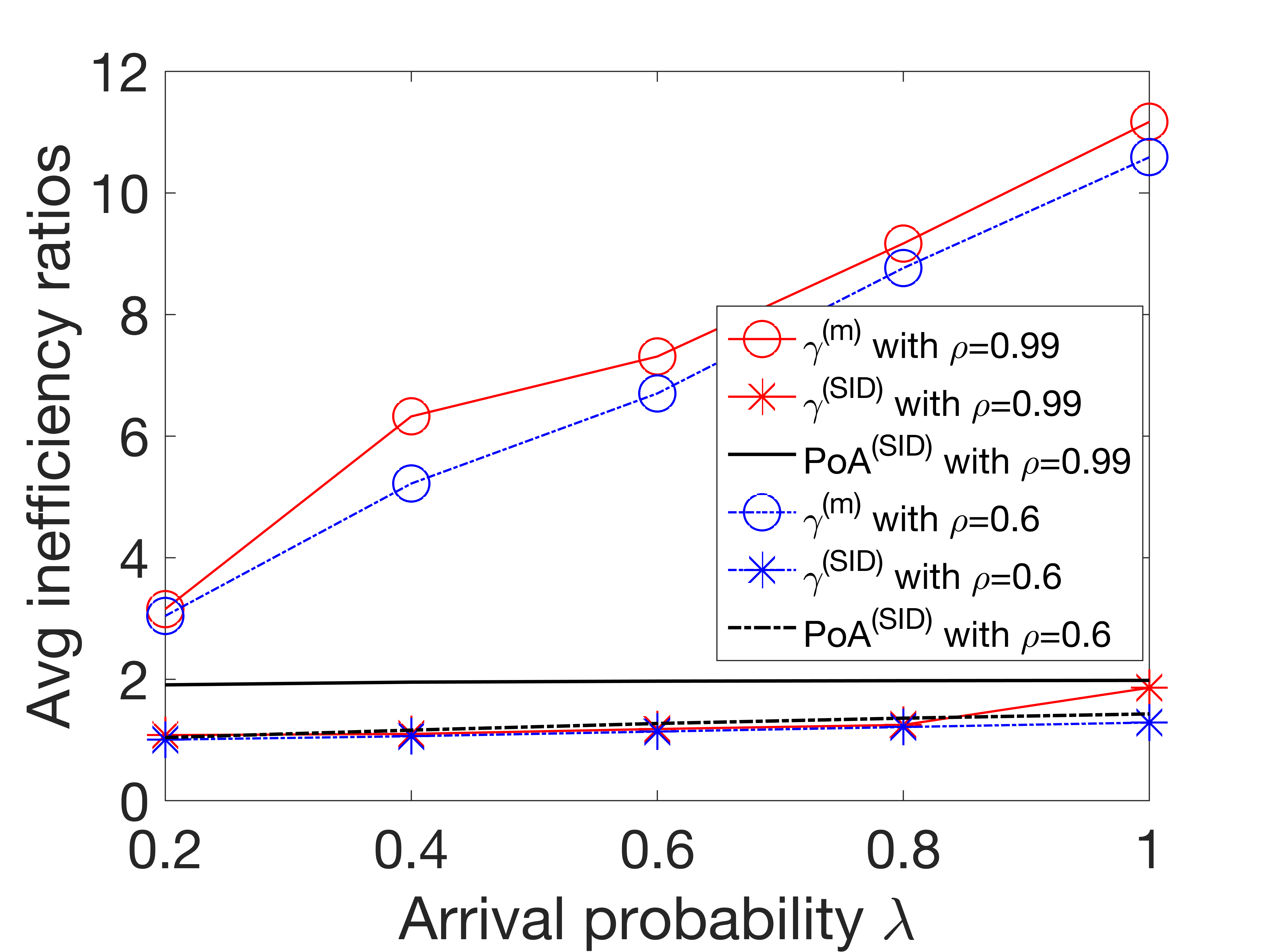}} 
    \caption{Average inefficiency ratios $\gamma^{(m)}$ under myopic policy in (\ref{cost_Cm}), $\gamma^{(\emptyset)}$ under hiding policy in (\ref{pi^empty}), and $\gamma^{(\text{SID})}$ under our SID mechanism.}
    \label{fig:avg_mechanism_comparison}
\end{figure}

In real-world scenarios, some users may gather additional traffic information from multiple platforms (e.g., Google Maps, Waze, Apple Maps), which could compromise the effectiveness of information hiding. However, in practice, not all users simultaneously use multiple platforms due to the extra management effort involved.  
As a result, some users rely solely on a single platform (i.e., SID mechanism platform here) for their path routing. 
Let $\phi\in(0,1)$ represent the probability that a user will only use a single source for information. 
In the following corollary, we examine whether our SID mechanism remains effective when users can access other information sources.
\begin{corollary}\label{corol:single_source}
If a portion $\phi\in(0,1)$ of users rely on a single source for information, our SID mechanism results in $\text{PoA}^{(\text{SID})} \leq \max\Big\{\frac{1}{1-\frac{\rho^{\frac{1}{\lambda}}}{2}}, \frac{1}{1-(1-\phi)\rho^{\frac{1}{\lambda}}}\Big\}$.  
\end{corollary}
The proof of Corollary~\ref{corol:single_source} is given in Appendix~H. Intuitively, when $\phi$ is close to $1$, most users will follow our SID mechanism to make path decisions, resulting in the same PoA as in Theorem~\ref{thm:poa_incentive}. Conversely, as $\phi \rightarrow 0$, most users will follow the myopic policy for path decisions, with only a small portion adhering to the SID mechanism, leading to a PoA similar to that in Proposition~\ref{thm:poa}.

Besides the worst-case performance analysis, we further verify our mechanism's average performance using extensive simulations. Define the following average inefficiency ratio between expected social costs achieved by our SID mechanism and social optimum in (\ref{cost_C*}):
\begin{equation}\label{gamma_SID}
\gamma^{(\text{SID})}=\frac{\mathbb{E}\big[C^{(\text{SID})}\big(\mathbf{L}(t),\mathbf{x}(t),s(t)\big)\big]}{\mathbb{E}\big[C^{*}\big(\mathbf{L}(t),\mathbf{x}(t),s(t)\big)\big]}.
\end{equation}
To compare, we define $\gamma^{(m)}$ to be the average inefficiency ratio between social costs achieved by the myopic policy in (\ref{cost_Cm}) and socially optimal policy in (\ref{cost_C*}). 
We similarly define $\gamma^{(\emptyset)}$ to be the average inefficiency ratio caused by the hiding policy in (\ref{pi^empty}).
After running $50$ long-term experiments for averaging each ratio, we plot Fig.~\ref{fig:avg_ratio} and Fig.~\ref{fig:rate_comparison} to compare $\gamma^{(m)}$ and $\gamma^{(\emptyset)}$ to $\gamma^{(\text{SID})}$ versus risky path number $N$ and arrival probability $\lambda$, respectively. Here we set $\alpha=0.99,\alpha_L=0,\Delta \ell=1, p_H=0.8, p_L=0.2, q_{HH}=0.99,q_{LL}=0.99$. At initial time $t=0$, we let $\ell_0(0)=100,\ell_i(0)=105$ and $x_i(0)=0.5$ for any path.

Fig.~\ref{fig:avg_ratio} shows that our SID mechanism obviously reduces $\gamma^{(m)}>10$ and $\gamma^{(\emptyset)}>11$ to $\gamma^{(\text{SID})}<2$ at $N=2$, which is consistent with Proposition \ref{thm:poa}, Proposition~\ref{lemma:pi^empty}, and Theorem~\ref{thm:poa_incentive}. 
Fig.~\ref{fig:avg_ratio} also shows that the efficiency loss due to users' selfish routing decreases with $N$, as more choices of risky paths help negate the hazard risk at each path. 
Furthermore, the observed trend that efficiency loss decreases with $N$ reflects realistic scenarios where more available roads allow users to better distribute their selections, thereby reducing congestion and hazard risks.
In Fig.~\ref{fig:avg_ratio}, we also vary high-hazard state $\alpha_H$ to make a comparison, and we see that a larger $\alpha_H$ causes less efficiency loss due to users' reduced explorations to risky paths for both myopic and hiding policies.

In Fig.~\ref{fig:rate_comparison}, we fix $N=2$ and $\alpha_H=2$ to illustrate that both average inefficiency ratios $\gamma^{(m)}$ and $\gamma^{(\text{SID})}$ increase with arrival probability $\lambda$ and discount factor $\rho$, which is consistent with Proposition~\ref{thm:poa} and Theorem~\ref{thm:poa_incentive}. In the worst-case scenario with $\rho=0.99$ and $\lambda=1$, our SID mechanism reduces $\gamma^{(m)}>10$ to $\gamma^{\text{SID}}<2$, which aligns with the observations in Fig.~\ref{fig:avg_ratio}.
We also plot the upper bounds of $\text{PoA}^{(\text{SID})}$ derived in Theorem~\ref{thm:poa_incentive} to verify that the $\gamma^{(\text{SID})}$ is always bounded by $\text{PoA}^{(\text{SID})}$.
Overall, these experimental findings verify the robustness of our SID mechanism and its superior performance compared to the myopic policy. To further demonstrate its practical applicability, we will validate that our SID mechanism consistently outperforms the myopic policy by real-world experiments, as detailed in Section~\ref{section7}.


\section{Extensions to General Linear Path Graphs and Dynamic Markov Chains}\label{section6}

In this section, we extend our system model and analysis to more general transportation networks with multiple intermediate nodes and risky paths, and allow the static Markov chain in Fig.~\ref{fig:POMDP} on risky paths to become dynamic. In this generalized system model, we will first derive the new PoA lower bound for the myopic policy, which also depends on the maximum variation of stochastic transition probabilities in the dynamic Markov chain. After that, we will show that our SID mechanism still works for the generalized system with the same PoA upper bound as in Theorem \ref{thm:poa_incentive}. Finally, we will conduct experiments to examine the average system performance of our SID mechanism.

\renewcommand\thefigure{\arabic{figure}(a)} 
\begin{figure*}[t]
    \centering
    \includegraphics[width=0.85\textwidth]{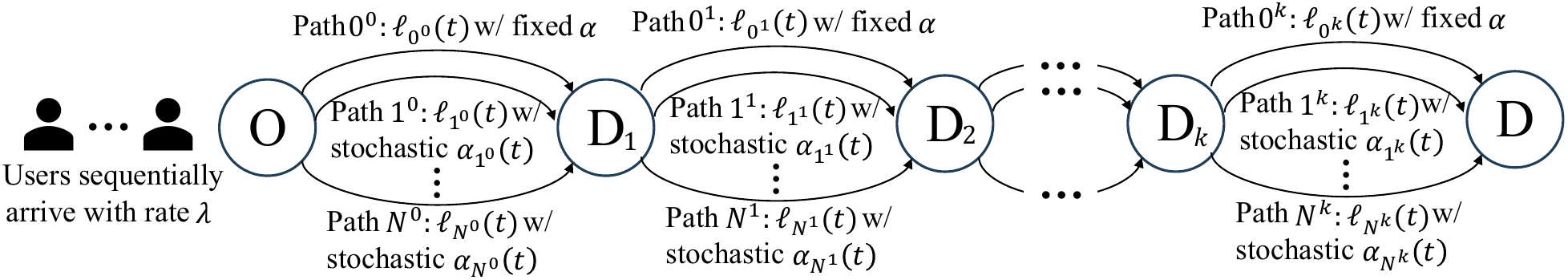}
    \captionsetup{font={footnotesize}} 
    \caption{Generalized linear path graph from parallel graph in Fig.~\ref{fig:congestion_game}: At the beginning of each time slot $t\in\{1,2,\cdots\}$, each user arrival at any node $\text{D}_j\in\{\text{O},\text{D}_1,\cdots, \text{D}_k\}$ selects one path from multiple available paths to travel to the next node in this linear path graph. Among all the available paths between intermediate nodes $\text{D}_{j}$ and $\text{D}_{j+1}$, where $j\in\{0,\cdots,k\}$, path $0^j$ is considered safe, while path $i^j\in\{1^j,\cdots,N^j\}$ is risky.}
    \label{fig:congestion_game_k}
\end{figure*}

\subsection{Extensions of System Model}
Before generalizing the parallel transportation network of our system model in Fig. \ref{fig:congestion_game}, we first introduce the definition of a widely used path graph in the following.
\begin{definition}[Linear Path graph \cite{Broersma1989pathgraph}]
A linear path graph has an ordered list of vertices, where an edge joining pairs of vertices represent a path. A path graph has two terminal nodes and multiple intermediate nodes.
\end{definition}

As illustrated in Fig. \ref{fig:congestion_game_k}, we model a general linear path graph $k$ of intermediate nodes, denoted by $\text{D}_1,\cdots, \text{D}_k$, lying between the origin O and the destination D. At the two ends, we let $\text{D}_0=\text{O}$ and $\text{D}_{k+1}=\text{D}$. At each node $\text{D}_{j}$ for $j\in\{0,1,\cdots,k\}$, there exist $N$ risky paths $1^{j},\cdots, N^{j}$ and one safe path $0^{j}$ leading to the subsequent node $\text{D}_{j+1}$. In this generalized network, users traveling on risky paths within each segment learn and update the actual traffic information there to the crowdsourcing platform.

Between any two adjacent nodes $\text{D}_{j}$ and $\text{D}_{j+1}$, where $j\in\{0,1,\cdots,k\}$, the safe path~$0^j$ has a fixed traffic coefficient $\alpha$. For the other $N^j$ risky paths, as illustrated in Fig.~\ref{fig:markov_chain_noise}, their coefficients alternate between a high-hazard state $\alpha_H$ and a low-hazard state $\alpha_L$ over time. 
Generalized from the static Markov chain shown in Fig.~\ref{fig:POMDP}, transition probabilities (i.e., $q_{LL}(t),q_{LH}(t),q_{HL}(t)$ and $q_{HH}(t)$) in Fig.~\ref{fig:markov_chain_noise} now follow random distributions to fluctuate over time. For instance, $q_{LL}(t)$ varies randomly within the range of $[\max\{0,q_L-\sigma\},\min\{1,q_L+\sigma\}]$, where $q_L$ is the mean of $q_{LL}(t)$ and $\sigma$ characterizes the maximum variation  from this expected value. Likewise, we denote the expected value of $q_{HH}(t)$ by $q_H$, and $q_{HH}(t)$ varies within $[\max\{0,q_H-\sigma\},\min\{1,q_H+\sigma\}]$.

\renewcommand\thefigure{4(b)} 
\begin{figure}[t]
    \centering
    \captionsetup{font={footnotesize}} \includegraphics[width=0.43\textwidth]{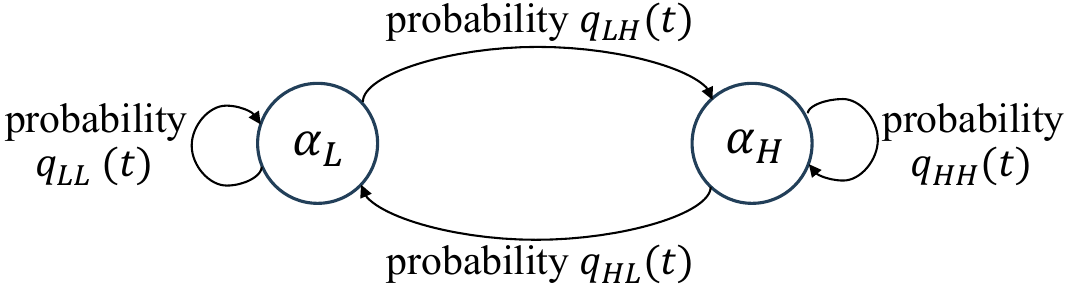}
    \caption{The dynamic Markov chain for modelling $\alpha_i(t)$ dynamics of stochastic path $i^j\in\{1^j,\cdots,N^j\}$. Unlike Fig. \ref{fig:POMDP}, here we relax to dynamic transition probabilities $q_{LL}(t),q_{LH}(t),q_{HL}(t)$ and $q_{HH}(t)$.}
    \label{fig:markov_chain_noise}
\end{figure}

Under the dynamic Markov chain extension in Fig.~\ref{fig:markov_chain_noise}, the update of hazard belief $x_{i^j}(t+1)$ of path $i^j$ in (\ref{x_i(t)}) becomes:
\begin{align}
    x_{i^j}(t+1)=x'_{i^j}(t)q_{HH}(t)+(1-x'_{i^j}(t))q_{LH}(t).\label{x_i^j(t+1)}
\end{align}

\subsection{Analysis of Myopic and Socially Optimal Policies}
Based on the estimated hazard belief $x_{i^j}(t)$ for each path $i^j\in\{1^j,\cdots,N^j\}$, users determine their routing choices when arriving at node $\text{D}_j$.
Although there are a total of $k+1$ road segments from O to D, users following either the myopic or the socially optimal policy only need to sequentially make routing decisions between nodes $\text{D}_j$ and $\text{D}_{j+1}$ upon arrival at $\text{D}_j$ as in Fig. \ref{fig:congestion_game}. Therefore, we can similarly formulate the long-term cost functions under both myopic and socially optimal policies for each node $\text{D}_j$ as (\ref{cost_Cm}) and (\ref{cost_C*}), respectively.

Before analyzing the PoA caused by the myopic policy, we similarly derive the stationary belief $\Bar{x}$ of each risky path:
\begin{align}\label{bar_x_noise}
    \Bar{x}=\mathbb{E}\left[\frac{1-q_{LL}(t)}{2-q_{LL}(t)-q_{HH}(t)}\right],
\end{align}
which is different from $\Bar{x}$ in (\ref{x_bar}) under the static Markov chain and depends on the distributions of $q_{LL}(t)$ and $q_{HH}(t)$. 
Given the new steady-state $\Bar{x}$ in (\ref{bar_x_noise}) for the general linear graph in Fig.~\ref{fig:congestion_game_k}, there still exists a belief threshold $x^{th}$ satisfying (\ref{x_th}) for each path $i^j$ as in Proposition~\ref{prop:explore}, which analytically tells the myopic policy misses both exploitation and exploration on risky paths over time. Then we analyze the PoA caused by the myopic policy in the following proposition.
\begin{proposition}\label{prop:poa_c}
Under the general linear path network in Fig.~\ref{fig:congestion_game_k} with the dynamic Markov chain in Fig. \ref{fig:markov_chain_noise}, as compared to the social optimum, the myopic policy achieves
\begin{align}
     \text{PoA}^{(m)}\geq \frac{1-\sigma\rho^{\frac{1}{\lambda}}}{1-\rho^{\frac{1}{\lambda}}},\label{PoA_multi_random}
\end{align}
which approaches infinity as $\rho\rightarrow 1$ and $\sigma\rightarrow 0$.
\end{proposition}

The proof of Proposition \ref{prop:poa_c} is given in Appendix~I. Under the linear path graph in Fig. \ref{fig:congestion_game_k}, both myopic and socially optimal policies just repeat decision-making for $k+1$ times from O to D. Yet the dynamics of Markov chains in Fig.~\ref{fig:markov_chain_noise} change the prior PoA ratio in Proposition~\ref{thm:poa} to (\ref{PoA_multi_random}). To understand this, we consider the same zero-exploration case with $q_L\rightarrow 1$ and $\alpha_L=0$ on a single segment $\{\text{D}_j,\text{D}_{j+1}\}$ as Proposition~\ref{thm:poa} to analyze PoA.
In this worst-case example, myopic users always choose safe path $0^j$, still resulting in the social cost $\frac{\ell_{0^j}(0)}{1-\rho^{\frac{1}{\lambda}}}$ for the segment $\{\text{D}_j,\text{D}_{j+1}\}$.
While the socially optimal policy requires the first user arrival to explore path $1^j$ to find possible $\alpha_L=0$ there. As $q_L\rightarrow 1$, path $1^j$ is expected to keep at the low-hazard state all the time, and the total travel cost since the next user is greatly reduced. 
However, given the maximum variation $\sigma$ for the dynamic transition probability $q_{LL}(t)$, there is a maximum probability $\sigma$ of switching to the high-hazard state for path~$1^j$. Therefore, the social cost under the optimal policy becomes no higher than \[\rho^{\frac{1}{\lambda}}\ell_{1^j}(0)+\frac{(1-\sigma)\rho^{\frac{2}{\lambda}}}{1-(1-\sigma)\rho^{\frac{1}{\lambda}}}\Delta\ell+\frac{\sigma\rho^{\frac{2}{\lambda}}\ell_{0^j}(0)}{1-\sigma\rho^{\frac{1}{\lambda}}}.\] Letting $\frac{\Delta \ell}{\ell_{0^j}(0)}\rightarrow 0$, we obtain PoA in (\ref{PoA_multi_random}). 

By observing (\ref{PoA_multi_random}), $\text{PoA}^{(m)}$ decreases with the maximum variation $\sigma$ for telling a smaller efficiency loss, as $\sigma$ increases the social cost under the socially optimal policy while does not influence that under the myopic policy. If $\sigma=0$, PoA in (\ref{PoA_multi_random}) becomes the maximum $\frac{1}{1-\rho^{\frac{1}{\lambda}}}$, which equals the original PoA in Proposition \ref{thm:poa}. 
As $\rho\rightarrow 1$ and $\sigma\rightarrow 0$, $\text{PoA}^{(m)}$ in (\ref{PoA_multi_random}) approaches infinity, which still requires our SID mechanism to regulate.
Note that if $\sigma\rightarrow 1$, the lower bound of $\text{PoA}^{(m)}$ in (\ref{PoA_multi_random}) approaches $1$, since the transition probabilities $q_{HH}(t)$ and $q_{LL}(t)$ vary within the widest possible range $[0,1]$. This makes the Markov chain highly unpredictable, affecting both the socially optimal policy and the myopic policy.

\subsection{SID Mechanism Design and Analysis} 

Based on our analysis of this general system model, we only need to focus on a single segment $\{\text{D}_j,\text{D}_{j+1}\}$ to regulate each user when he arrives at node $\text{D}_j$. As a result, our SID mechanism in Definition \ref{def:mechanism} still works. In the next proposition, we prove the PoA upper bound of our SID mechanism.
\begin{proposition}\label{prop:PoA_incentive_multi}
Under the general linear path network in Fig.~\ref{fig:congestion_game_k} with the dynamic Markov chain in Fig. \ref{fig:markov_chain_noise}, our SID mechanism reduces PoA in (\ref{PoA_multi_random}) to $\text{PoA}^{(\text{SID})}\leq\frac{1}{1-\frac{\rho^{\frac{1}{\lambda}}}{2}}$, which is always no more than $2$.
\end{proposition}
The proof of Proposition \ref{prop:PoA_incentive_multi} is similar to Theorem \ref{thm:poa_incentive} and is given in Appendix J. Based on our analysis of Proposition \ref{prop:poa_c}, the system performs the worst under $\sigma=0$. Therefore, we consider the same maximum-exploration case as Theorem \ref{thm:poa_incentive} to derive this PoA upper bound.

\setcounter{figure}{4} 
\renewcommand\thefigure{\arabic{figure}} 

\begin{figure}[t]
    \centering
    \captionsetup{font={footnotesize}}
    \subfigure[We set $q_{H}=0.9$ and $q_{L}=0.99$, such that $\Bar{x}$ in (\ref{bar_x_noise}) increases with $\sigma$.]{\label{fig:noise_comparison2}\includegraphics[width=0.35\textwidth]{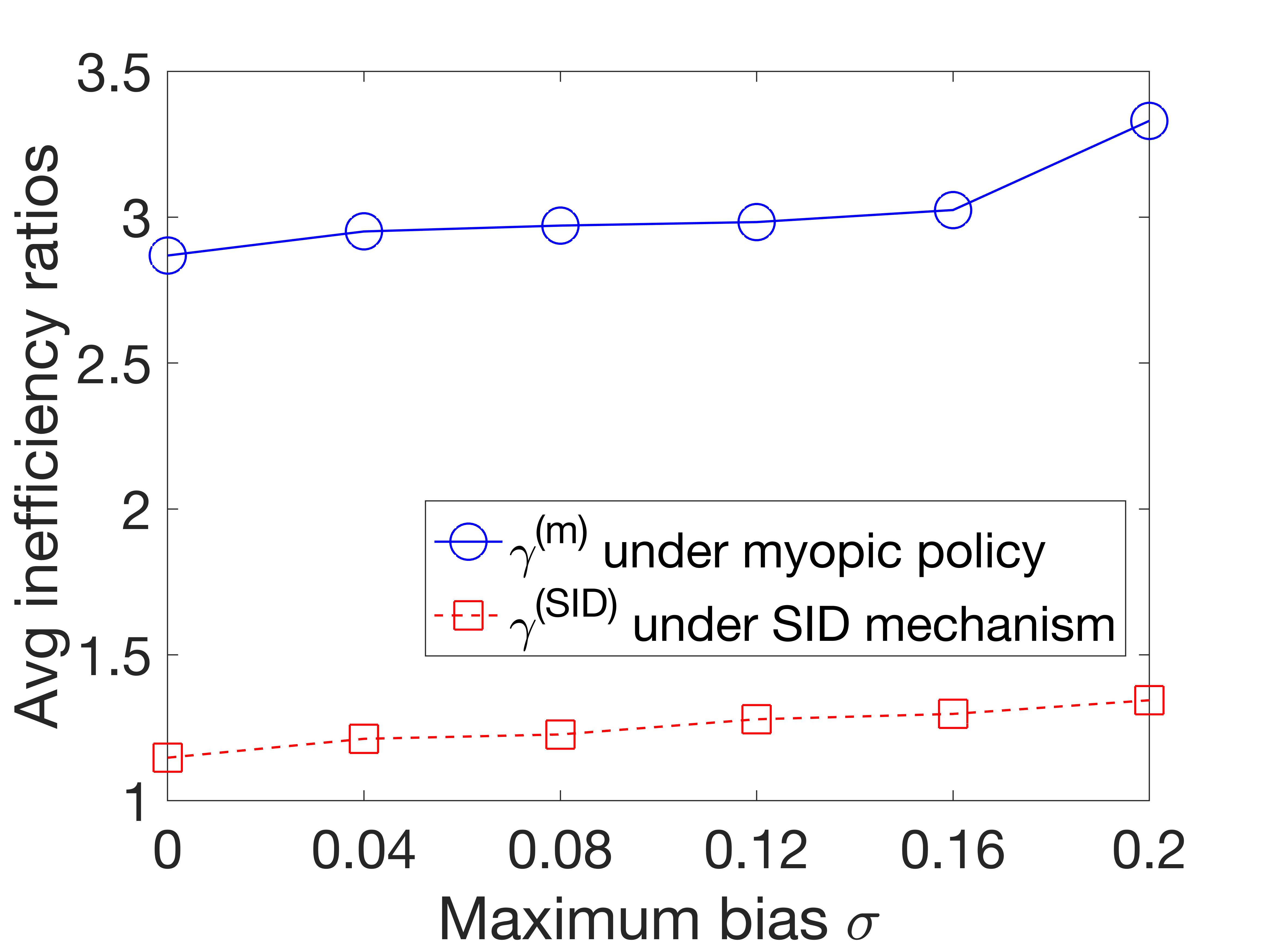}} 
    \subfigure[We set $q_{H}=0.99$ and $q_{L}=0.9$, such that $\Bar{x}$ in (\ref{bar_x_noise}) decreases with $\sigma$.]{\label{fig:noise_comparison1}\includegraphics[width=0.35\textwidth]{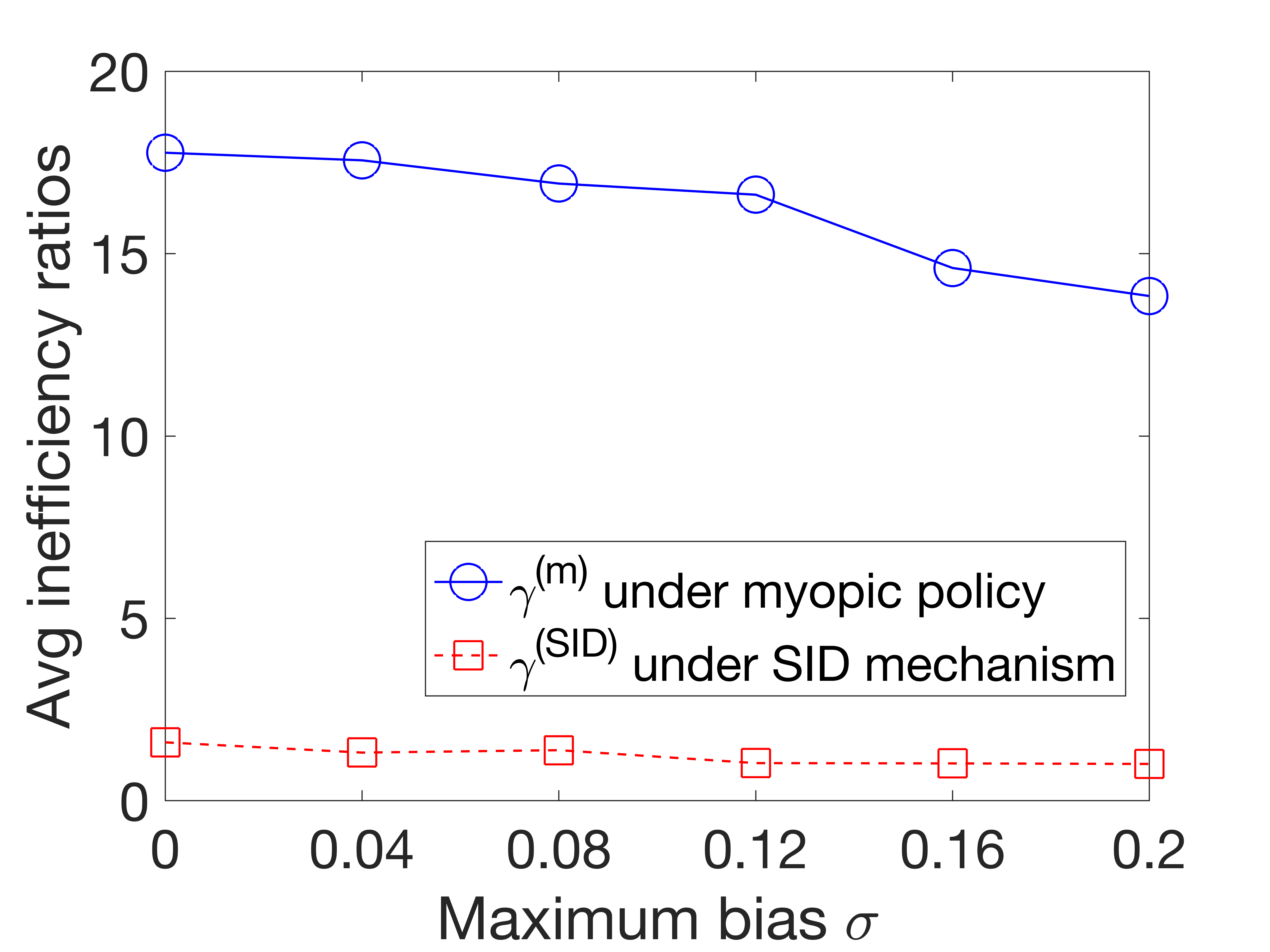}} 
    \caption{Average inefficiency ratios $\gamma^{(m)}$ under the myopic policy and $\gamma^{(\text{SID})}$ under our SID mechanism in the linear path graph. We vary the maximum variation  $\sigma$ of transition probabilities in set $\{0,0.04,0.08,0.12,1.16,0.20\}$. Here $q_{HH}(t)$ and $q_{LL}(t)$ satisfy uniform distributions on intervals $[\max\{0,q_H-\sigma\},\min\{1,q_H+\sigma\}]$ and $[\max\{0,q_L-\sigma\},\min\{1,q_L+\sigma\}]$, respectively. }
    \label{fig:noise_comparison}
\end{figure}

Besides the PoA analysis, to tell and explain the impact of $\sigma$, we further conduct numerical experiments to examine the average performance ratio of $\gamma^{(\text{SID})}$ under our SID mechanism in (\ref{gamma_SID}), as compared to $\gamma^{(m)}$ under the myopic policy. We set $\lambda=1, \alpha=0.99,\alpha_L=0,\alpha_H=2,\Delta \ell=1, p_H=0.8, p_L=0.2,k=3$ for the dynamic system. At initial time $t=0$, we let $\ell_0(0)=100,\ell_i(0)=105$ and $x_i(0)=0.5$ for any path~$i$. 
Here we let $q_{HH}(t)$ and $q_{LL}(t)$ satisfy uniform distributions independently in each time slot. 
In Fig. \ref{fig:noise_comparison}, we present two sets of parameters $q_{L}$ and $q_H$, which leads to different monotonicity patterns of $\Bar{x}$ in (\ref{bar_x_noise}) with respect to $\sigma$,
to illustrate the impact of $\sigma\in\{0,0.04,0.08,0.12,0.16,0.2\}$ on the average inefficiency ratios $\gamma^{(m)}$ and $\gamma^{(\text{SID})}$, respectively. In both cases, our SID mechanism leads to obviously smaller efficiency loss as compared to the myopic policy. 

In Fig.~\ref{fig:noise_comparison2}, we set $q_{H}=0.9$ and $q_L=0.99$, such that $\Bar{x}$ in (\ref{bar_x_noise}) increases with $\sigma$. In this case, both $\gamma^{(\text{SID})}$ and $\gamma^{(m)}$ increase with maximum probability variation $\sigma$ of the Markov chains. This is because as $\Bar{x}$ increases with $\sigma$, risky paths with higher $\Bar{x}$ become more congested, which results in increased $\gamma^{(m)}$. At the same time, our SID mechanism in Definition~\ref{def:mechanism} discloses system information to more user arrivals to encourage these users to follow the myopic policy. Thus, this increases the efficiency loss for $\gamma^{(\text{SID})}$.

In Fig.~\ref{fig:noise_comparison1}, we set $q_{H}=0.99$ and $q_L=0.9$, such that $\Bar{x}$ in (\ref{bar_x_noise}) decreases with $\sigma$. We observe that $\gamma^{(m)}$ decreases from $18$ at $\sigma=0$ to less than $15$ at $\sigma=0.2$, and our $\gamma^{(\text{SID})}$ reduces and approaches to the optimum. This is because as $\Bar{x}$ decreases with $\sigma$, a smaller steady hazard belief $\Bar{x}$ results in reduced congestion on risky paths, thus lowering the social cost for the myopic policy. At the same time, our SID hides actual system information from more users, and there are more users following our SID mechanism's optimal routing recommendations. Thus, this reduces the efficiency loss for~$\gamma^{(\text{SID})}$.



\section{Experiment Validation Using Real Datasets}\label{section7}

Besides the worst-case analysis in the last section, we further conduct experiments using real datasets to evaluate our SID mechanism's average performance versus the myopic policy and the socially optimal policy. To further practicalize our congestion model in (\ref{L_i(t+1)}), we sample peak hours' real-time traffic congestion data of a linear path graph network in Shanghai, China on weekends using BaiduMap dataset \cite{Baidumap}. 

\begin{figure}[t]
    \centering
    \captionsetup{font={footnotesize}}
    \includegraphics[width=0.45\textwidth]{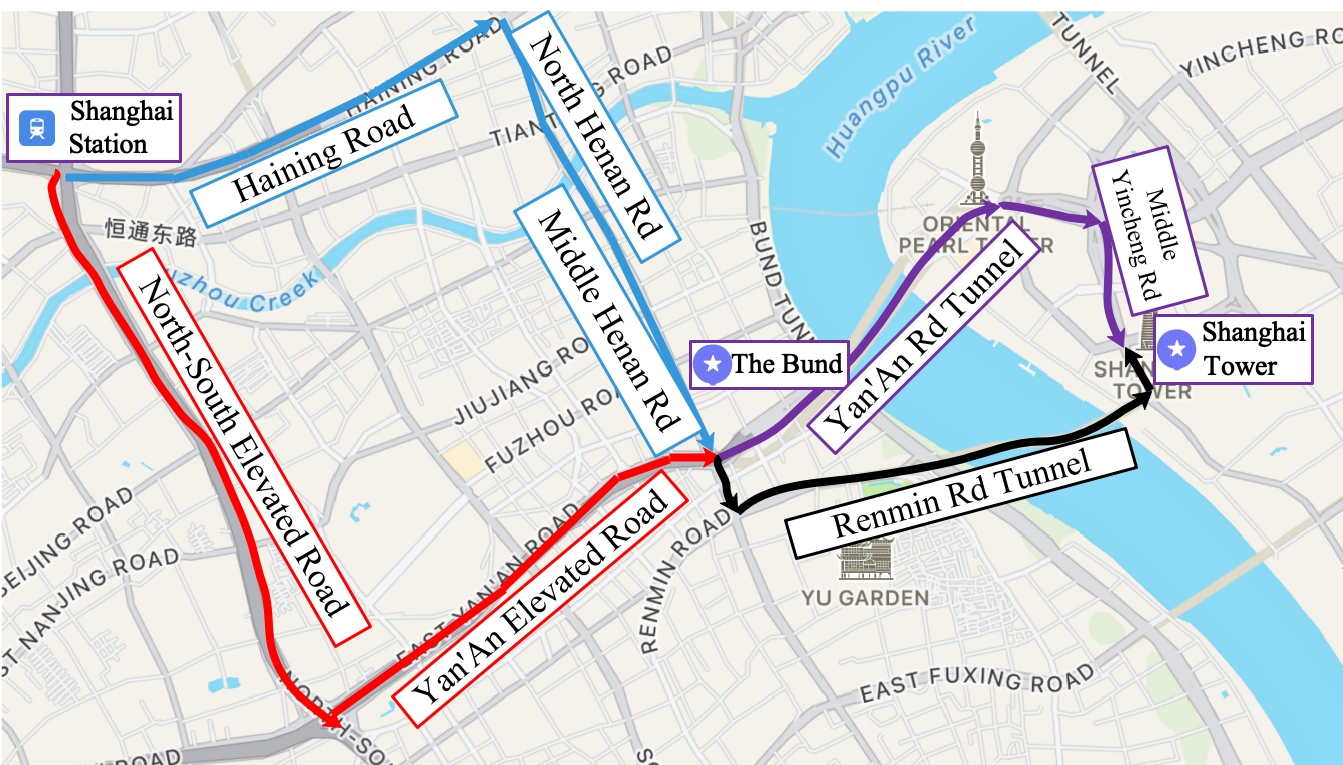}
    \caption{A typical linear path network from Shanghai Station to Shanghai Tower, passing through the Bund as an intermediate node, includes several route options. From Shanghai Station to the Bund, travelers can choose between Haining Road, North Henan Road, and Middle Henan Road (in blue), or North-South Elevated Road and Yan'An Elevated Road (in red). From the Bund to Shanghai Tower, the options include Yan'An Road Tunnel and Middle Yincheng Road (in purple), or Renmin Road Tunnel (in black).}
    \label{fig:Shanghai_map}
\end{figure}

In Fig. \ref{fig:Shanghai_map}, we illustrate a popular linear path network from Shanghai Station to Shanghai Tower during weekends, passing through the Bund as an intermediate node. This reflects the common flow of visitors who leave Shanghai Station and travel to the Bund and Shanghai Tower. From Shanghai Station to the Bund, travelers have the following route options:
\begin{itemize}
    \item Path $1^0$: First Haining Road, then North Henan Road, and finally Middle Henan Road.
    \item Path $2^0$: First North-South Elevated Road, and then Yan'An Elevated Road.
\end{itemize}
From the Bund to Shanghai Tower, travelers can choose between:
\begin{itemize}
    \item Path $1^1$: First Yan'An Elevated Road, and then Middle Yincheng Road.
    \item Path $2^1$: Renmin Road Tunnel.
\end{itemize}

We sampled $728$ statuses for the $8$ road segments, with data collected every $2$ minutes over a $3$-hour period. We validate using the dataset that the traffic conditions on North-South Elevated Road and Yan'An Elevated Road in path segment $2^0$, as well as Yan'An Road Tunnel and Middle Yincheng Road in path segment $1^1$, can be effectively approximated as Markov chains with two discretized states (high and low traffic states) as in Fig.~\ref{fig:POMDP}. In contrast, Haining Road, North Henan Road, and Middle Henan Road in path segment $1^0$, as well as Renmin Road Tunnel on path segment $2^1$, tend to exhibit deterministic/safe conditions. Similar to \cite{li2024optimize,eddy1998profile,chen2016predicting}, we employ the hidden Markov model (HMM) approach to train the transition probability matrices for the four stochastic road segments. 
Below, we present our obtained average transition matrices NS\_E, YA\_E, YA\_T, M\_YC for North-South Elevated Road, Yan'An Elevated Road, Yan'An Road Tunnel, and Middle Yincheng Road respectively:
\begin{align*}
    &\text{NS\_E}=\begin{bmatrix}
        0.8947 & 0.1053\\
        0.1000 & 0.9000
    \end{bmatrix},\text{YA\_E}=\begin{bmatrix}
        0.7692 & 0.2308\\
        0.2500 & 0.7500
    \end{bmatrix}.\\
    &\text{YA\_T}=\begin{bmatrix}
        0.8213 & 0.1787\\
        0.1490 & 0.8510
    \end{bmatrix},
    \text{M\_YC}=\begin{bmatrix}
        0.6387 & 0.3613\\
        0.3578 & 0.6422
    \end{bmatrix}.
\end{align*}

Under the myopic policy, each selfish user assesses potential path choices by summing the costs of individual road segments (e.g., Haining Road, North Henan Road, and Middle Henan Road for path segment $1^0$) to select the path segment that minimizes his own travel cost. In contrast, the socially optimal policy aims to identify the path choice that minimizes the aggregate long-term social cost for all users. Our SID mechanism selectively hides or discloses information about each road to users, depending on the actual hazard belief of each stochastic road segment. As users’ costs are proportional to their travel delays, we conduct $100$ experiments by using the real data on Fig.~\ref{fig:Shanghai_map}, each with $101$ time slots (equivalent to $202$ minutes), to calculate their average long-term social costs (in minutes).
Based on the travel latency data, we calculate the long-term average correlation coefficients $\alpha_H\approx 1.5$, $\alpha_L\approx 0.3, \alpha\approx 0.6$ based on the linear travel latency functions in (\ref{L_0(t+1)}) and (\ref{L_i(t+1)}). At initial time, we observe the travel latencies (in minutes) in each path segment as $\ell_{1^0}(0)=25$, $\ell_{2^0}(0)=29,\ell_{1^1}(0)=12,\ell_{2^1}(0)=9$. Here we set average arrival probability $\lambda=0.95$ and $\rho=0.95$.

\begin{figure}[t]
    \centering
    \captionsetup{font={footnotesize}}
    \includegraphics[width=0.35\textwidth]{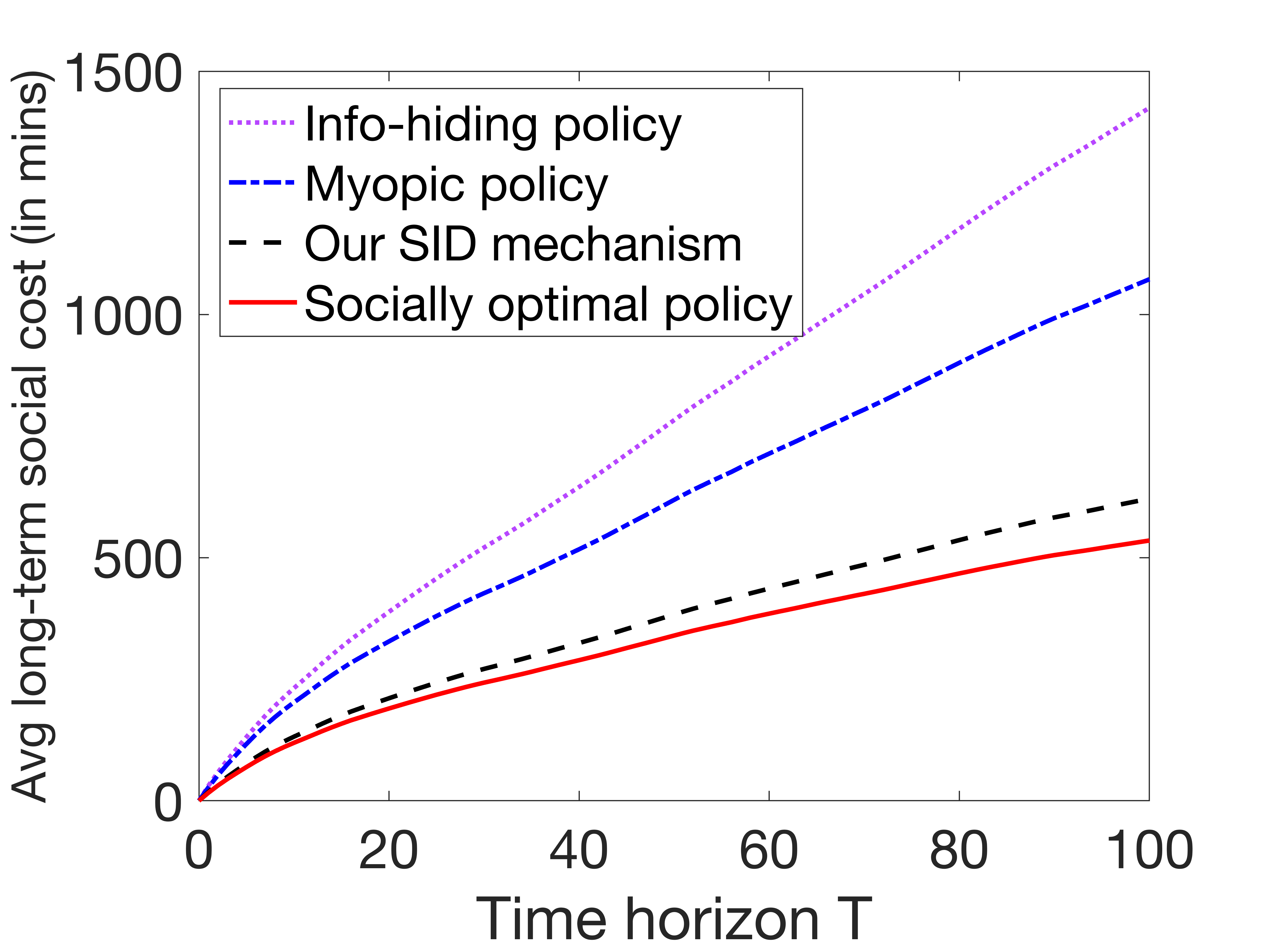}
    \caption{Average long-term costs (in minutes) under information-hiding, myopic, socially optimal policies, and our SID versus time horizon $T$.}
    \label{fig:Map_comparison}
\end{figure}

Fig. \ref{fig:Map_comparison} compares the average long-term social costs under information-hiding, myopic, socially optimal policies, and our SID mechanism versus time horizon $T$. It depicts that our SID mechanism incurred no more than $20\%$ social cost from the optimum for any time horizon $T$, while the existing myopic policy (used by Google Maps and Waze) and the information-hiding policy (used in existing literature \cite{tavafoghi2017informational} and \cite{li2019recommending}) cause more than $100\%$ and $180\%$ efficiency loss, respectively.

\section{Conclusion}
Our work is the first to leverage information learning to improve system performance in congestion games, and our proposed SID mechanism provides a practical solution for incentivizing users to adopt better routing policies. 
First, we consider a simple but fundamental parallel routing network with one deterministic path and multiple stochastic paths for users with an average arrival probability $\lambda$. We prove that the current myopic routing policy (widely used in Waze and Google Maps) misses both exploration (when strong hazard belief) and exploitation (when weak hazard belief) as compared to the social optimum. Due to the myopic policy's under-exploration, we prove that the caused price of anarchy (PoA) is larger than \(\frac{1}{1-\rho^{\frac{1}{\lambda}}}\), which can be arbitrarily large as discount factor \(\rho\rightarrow1\). 
To mitigate such huge efficiency loss, we propose a novel selective information disclosure (SID) mechanism: we only reveal the latest traffic information to users when they intend to over-explore stochastic paths upon arrival, while hiding such information when they want to under-explore. We prove that our mechanism successfully reduces PoA to be less than~\(2\). 
Besides the parallel routing network, we further extend our mechanism and PoA results to any linear path graphs with multiple intermediate nodes. 
In addition to the worst-case performance evaluation, we conduct extensive simulations with both synthetic and real transportation datasets to demonstrate the close-to-optimal average-case performance of our SID mechanism.

Several potential future directions could extend this work. For instance, analyzing the causes of congestion could lead to better estimates of the latency set $\mathbf{L}(t)$ and hazard belief set $\mathbf{x}(t)$ for both the myopic and the socially optimal policy. 
Additionally, considering scenarios with multiple simultaneous user arrivals could help balance the number of users selecting each risky path. 
Furthermore, we plan to extend our analysis to more complex transportation networks, such as those with multiple destination nodes, where accounting for network topology would be crucial for optimizing users' routing decisions.

\begin{IEEEbiography}[{\includegraphics[width=1in,height=1.25in,clip,keepaspectratio]{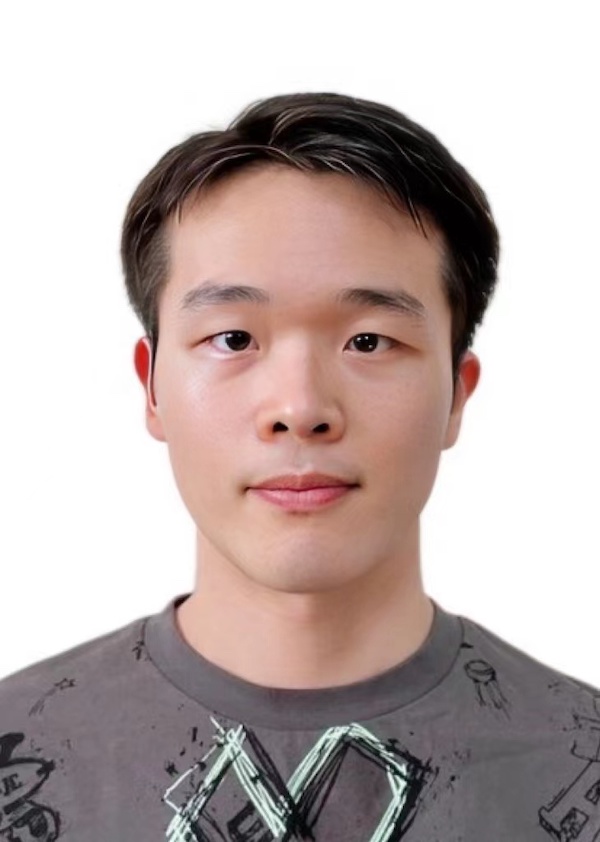}}]{Hongbo Li}(S'24-M'24)
received the Ph.D. degree from Singapore University of Technology and Design (SUTD) in 2024. He is currently a Postdoctoral Research Fellow with the Pillar of Engineering Systems and Design, SUTD. In 2024, he was a Visiting Scholar at The Ohio State University (OSU), Columbus, OH, USA. His research interests include networked AI, machine learning theory, game theory and mechanism design.
\end{IEEEbiography}

\vspace{11pt}
\begin{IEEEbiography}[{\includegraphics[width=1in,height=1.25in,clip,keepaspectratio]{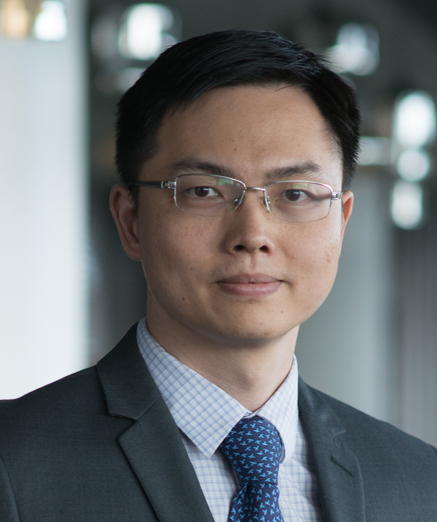}}]{Lingjie Duan}(S'09-M'12-SM'17) received the Ph.D. degree from The Chinese University of Hong Kong in 2012. He is an Associate Professor at the Singapore University of Technology and Design (SUTD) and is an Associate Head of Pillar (AHOP) of Engineering Systems and Design. In 2011, he was a Visiting Scholar at University of California at Berkeley, Berkeley, CA, USA. His research interests include network economics and game theory, network security and privacy, energy harvesting wireless communications, and mobile crowdsourcing. He is an Associate Editor of IEEE/ACM Transactions on Networking and IEEE Transactions on Mobile Computing. He was an Editor of IEEE Transactions on Wireless Communications and IEEE Communications Surveys and Tutorials. He also served as a Guest Editor of the IEEE Journal on Selected Areas in Communications Special Issue on Human-in-the-Loop Mobile Networks, as well as IEEE Wireless Communications Magazine. He is a General Chair of WiOpt 2023 Conference and is a regular TPC member of some other top conferences (e.g., INFOCOM, MobiHoc, SECON). He received the SUTD Excellence in Research Award in 2016 and the 10th IEEE ComSoc Asia-Pacific Outstanding Young Researcher Award in 2015.
\end{IEEEbiography}

\newpage

\appendices
\section{Proof of Lemma 1}\label{proof:lemma1}
We only need to prove that $C^{(m)}\big(\mathbf{L}(t), \mathbf{x}(t),s(t)\big)$ under the myopic policy increases with any path's expected latency $\mathbb{E}[\ell_i(t)|x_i(t-1),y_i(t-1)]$ in $\mathbf{L}(t)$ and $\mathbf{x}(t)$. Then the monotonicity of $C^*\big(\mathbf{L}(t), \mathbf{x}(t),s(t)\big)$ under the socially optimal policy similarly holds.

First, we prove the monotonicity of $C^{(m)}\big(\mathbf{L}(t), \mathbf{x}(t),s(t)\big)$ with expected latency $\mathbf{L}(t)$, where we consider a simple two-path network with safe path 0 and risky path 1. In this network, let $\mathbf{L}_a(t)=\mathbb{E}[\ell_1(t)|x_i(t-1),y_i(t-1)]$ and $\mathbf{L}_b(t)=\mathbb{E}[\ell_1(t)+1|x_i(t-1),y_i(t-1)]$ denote the two different travel latency sets of risky path 1, respectively. We want to prove $C^{(m)}\big(\mathbf{L}_a(t), \mathbf{x}(t),s(t)\big)\leq C^{(m)}\big(\mathbf{L}_b(t), \mathbf{x}(t),s(t)\big)$ given $\mathbf{L}_a(t)<\mathbf{L}_b(t)$ at current $t$.
According to (15), the probabilities $\text{P}\big(y_{1}(t)=1|\mathbf{x}(t)\big)$ of observing a hazard on path $1$ under $\mathbf{L}_a(t)$ and $\mathbf{L}_b(t)$ are always the same. Hence, we have $\mathbf{L}_a(\tau)<\mathbf{L}_b(\tau)$ for any $\tau> t$ based on their dynamics in (9). While on safe path 0, the travel latency is always the same for both cost functions $C^{(m)}\big(\mathbf{L}_a(t), \mathbf{x}(t),s(t)\big)$ and $ C^{(m)}\big(\mathbf{L}_b(t), \mathbf{x}(t),s(t)\big)$. As a result, $C^{(m)}\big(\mathbf{L}_a(t), \mathbf{x}(t),s(t)\big)\leq C^{(m)}\big(\mathbf{L}_b(t), \mathbf{x}(t),s(t)\big)$ is always true, and this conclusion holds for the other transportation network with multiple risky paths.

Next, we prove that $C^{(m)}\big(\mathbf{L}(t), \mathbf{x}(t),s(t)\big)$ increases with hazard belief $\mathbf{x}(t)$. Since a larger $x_i(t)$ leads to a larger expected coefficient $\mathbb{E}[\alpha_i(t)|x'_i(t)]$ in (8), the future expected travel latency $\mathbb{E}[\ell_i(t)+1|x_i(t-1),y_i(t-1)]$ on risky path~$i$ increases with $x_i(t)$ based on (9). Based on the conclusion that $C^{(m)}\big(\mathbf{L}(t), \mathbf{x}(t),s(t)\big)$ increases with travel latency $\mathbb{E}[\ell_i(t)|x_i(t-1),y_i(t-1)]$, $C^{(m)}\big(\mathbf{L}(t), \mathbf{x}(t),s(t)\big)$ also increases with $\mathbf{x}(t)$.

We can use the same method to prove that the cost function $C^*\big(\mathbf{L}(t), \mathbf{x}(t),s(t)\big)$ under the socially optimal policy has the same monotonicity as $C^{(m)}\big(\mathbf{L}(t), \mathbf{x}(t),s(t)\big)$ under the myopic policy. This is straightforward as both cost functions hold similar dynamics. Note that one can also prove Lemma~1 using Bellman equation techniques and we skip it here.

\section{Proof of Proposition 1}\label{proof:prop1}
It is straightforward to derive the exploration thresholds $\ell^{(m)}(t)$ under the myopic policy and $\ell_i^*(t)$ under the socially optimal policy, thus we only prove that $\ell_i^*(t)$ increases with $x_i(t)$ in the following. For ease of exposition, we assume $\lambda=1$ in the following. Note that $\lambda$ does not influence the monotonicity of $\ell^{(m)}(t)$ and $\ell_i^*(t)$ in Proposition 1, as both policies hold the same arrival rate.

From the expression of $\ell_i^*(t)$ in (20), if we prove
\begin{align}
    \frac{\partial \Big(Q_0^*(t+1)-Q^*_{i}(t+1)\Big)}{\partial x_i(t)}\geq 0, \label{partial_x}
\end{align}
we can say that $\ell_i^*(t)$ increases with $x_i(t)$. Therefore, in the following, we first formulate $\frac{\partial Q_0^*(t+1)}{\partial x_{\hat{\iota}_1}(t)}$ in (\ref{partial_Q0}) and $\frac{\partial Q_{\hat{\iota}_1}^*(t+1)}{\partial x_{\hat{\iota}_1}(t)}$ in (\ref{partial_Qi}), respectively. Then we will apply mathematical induction to prove (\ref{partial_x}) based on the two formulations.

Denote the current optimal path by $\hat{\iota}_1$ for later use. Since the dynamics of travel latency on any path $i\neq \hat{\iota}_1$ is not related with $x_{\hat{\iota}_1}(t)$, $\frac{\partial \ell_i(t)}{\partial x_{\hat{\iota}_1}(t)}$ always equals 0. In consequence, according to the definitions of $Q_0^*(t+1)$ and $Q^*_{i}(t)$, we have
\begin{align}
    \frac{\partial Q_0^*(t+1)}{\partial x_{\hat{\iota}_1}(t)}=&\rho^{T_0}\frac{\partial \mathbb{E}_{T_0}\big[\ell_{\hat{\iota}_1}(t+T_0)|\pi^*(t)=0\big]}{\partial x_{\hat{\iota}_1}(t)}\notag\\&+\rho^{T_0+1} \frac{\partial Q_{\hat{\iota}_1}(t+T_0+1)}{\partial x_{\hat{\iota}_1}(t)},\label{partial_Q0}
\end{align}
where $T_0$ is the elapsed time slots until the next exploration to path $\hat{\iota}_1$ since time $t$, and $\mathbb{E}[\ell_{\hat{\iota}_1}(t+T_0)|\pi^*(t)=0]$ is the expected travel latency at $t+T_0$ conditional on current policy $\pi^*(t)=0$. Similarly, we can obtain 
\begin{align}
    \frac{\partial Q_{\hat{\iota}_1}^*(t+1)}{\partial x_{\hat{\iota}_1}(t)}=&\rho^{T_1}\frac{\partial \mathbb{E}_{T_1}\big[\ell_{\hat{\iota}_1}(t+T_1)|\pi^*(t)=\hat{\iota}_1\big]}{\partial x_{\hat{\iota}_1}(t)}\notag\\&+\rho^{T_1+1} \frac{\partial Q_{\hat{\iota}_1}(t+T_1+1)}{\partial x_{\hat{\iota}_1}(t)},\label{partial_Qi}
\end{align}
where $T_1$ is the elapsed time slots until the next exploration to path $\hat{\iota}_1$ for $Q_{\hat{\iota}_1}^*(t)$ after time $t$, and $\mathbb{E}[\ell_{\hat{\iota}_1}(t+T_1)|\pi^*(t)=\hat{\iota}_1]$ is the travel latency at $t+T_1$ conditional on $\pi^*(t)=\hat{\iota}_1$. Note that $T_0\leq T_1$ because the exploration to path $\hat{\iota}_1$ at time $t$ increases the travel latency by $\Delta \ell$, making latter users less willing to explore this path again. Based on formulations (\ref{partial_Q0}) and (\ref{partial_Qi}), we next use mathematical induction to prove (\ref{partial_x}).

First, if the time horizon $T=1$, (\ref{partial_x}) is obviously true because $T_0\leq T_1$. Note that if $T_0=T_1=1$, then $ \frac{\partial Q_0^*(t+1)}{\partial x_{\hat{\iota}_1}(t)}- \frac{\partial Q_{\hat{\iota}_1}^*(t+1)}{\partial x_{\hat{\iota}_1}(t)}=0$.
Next, we assume the induction hypothesis that (\ref{partial_x}) is still true for a larger time horizon $T$, where $T\gg T_1\geq T_0$. It follows to show that (\ref{partial_x}) is still true for time horizon $T+1$. Since $T\gg T_1\geq T_0$, we only need to compare $\rho^{T_0+1} \frac{\partial Q_{\hat{\iota}_1}(t+T_0+1)}{\partial x_{\hat{\iota}_1}(t)}$ in (\ref{partial_Q0}) to $\rho^{T_1+1} \frac{\partial Q_{\hat{\iota}_1}(t+T_1+1)}{\partial x_{\hat{\iota}_1}(t)}$ in (\ref{partial_Qi}). Since the left time slots for cost-to-go $Q_{\hat{\iota}_1}(t+T_0+1)$ and $Q_{\hat{\iota}_1}(t+T_1+1)$ are $T-T_0$ and $T-T_1$, respectively, we can obtain that 
\[\rho^{T_0+1} \frac{\partial Q_{\hat{\iota}_1}(t+T_0+1)}{\partial x_{\hat{\iota}_1}(t)}-\rho^{T_1+1} \frac{\partial Q_{\hat{\iota}_1}(t+T_1+1)}{\partial x_{\hat{\iota}_1}(t)}\geq 0\]
in (\ref{partial_Q0}) and (\ref{partial_Qi}) based on the conclusion that (\ref{partial_x}) is true when its time horizon is $T-T_0$.

As we have proven that both the base case and the induction step are true, (\ref{partial_x}) is true for each time horizon $T$. This completes the proof that $\ell_i^*(t)$ increases with $x_i(t)$.

\section{Proof of Lemma 2}\label{proof:lemma2}
As the stochastic path number $N$ increases, $\ell^*_i(t)$ and $\ell^{(m)}(t)$ will approach identical because more paths help negate the congestion. Hence, we consider the worst-case with the two-path transportation network to prove (24), then (24) is always true for the multi-path network with $N\geq 2$. To prove (24), we need to prove that the exploration loss $\Delta C_1$ for the current user under the socially optimal policy is always smaller than the long-term benefit $\Delta C_2$ for future users under the socially optimal policy, as compared to the myopic policy.

Take a system with $\pi^*(t)=1$ under the socially optimal policy and $\pi^{(m)}(t)=2$ under the myopic policy as an example, where $\ell_0(t)$ is no less than $\mathbb{E}[\ell_1(t)|x_1(t-1),y_1(t-1)]$. Define $\Delta C_1$ to be the extra travel latency of choosing path 0 instead of path 1 for the current user, which is
\begin{align*}
    \Delta C_1=&\mathbb{E}[\ell_1(t)|x_1(t-1),y_1(t-1)]\\& -\mathbb{E}[\ell_2(t)|x_2(t-1),y_2(t-1)].
\end{align*}
Note that $\Delta C_1$ is the exploration cost for the current user caused by the socially optimal policy.

We further define $\Delta C_2$ to be the exploitation benefit for latter users due to the exploration of path 1 under the socially optimal policy, which has the following upper bound
\begin{align*}
    \Delta C_2 \leq& \rho \mathbb{E}[\ell_1(t)|x_1(t-1),y_1(t-1)]\\&+\rho^2 \mathbb{E}[\ell_1(t)|x_1(t-1),y_1(t-1)]+\cdots\\
    =& \frac{\rho\mathbb{E}[\ell_1(t)|x_1(t-1),y_1(t-1)]}{1-\rho}.
\end{align*}
This is because the travel latency for each user after time $t$ can be reduced at most $\mathbb{E}[\ell_1(t)|x_1(t-1),y_1(t-1)]$ if there is always low-hazard state $\alpha_L=0$ for any latter users exploring risky path 1. To make sure that $\pi^*(t)=0$ leads to the minimal social cost, we need
\begin{equation*}
    \Delta C_1 \leq \Delta C_2,
\end{equation*}
such that the long-term benefit can negate the current extra travel cost and the current policy $\pi^*(t)=1$ is the social optimum. By solving the above equality, we finally obtain
\begin{equation*}
    \mathbb{E}[\ell_1(t)|x_1(t-1),y_1(t-1)]\leq \frac{1}{1-\rho} \mathbb{E}[\ell_2(t)|x_2(t-1),y_2(t-1)],
\end{equation*}
which can be generalized to (24).

\section{Proof of Proposition 2}\label{proof:prop2}
To prove Proposition 2, we only need to prove that the myopic policy over-explores stochastic path $i$ (i.e., $\ell_i^*(t)\leq \ell^{(m)}(t)$) if $x_i(t)<\min\Big\{\frac{\alpha-\alpha_L}{\alpha_H-\alpha_L},\Bar{x}\Big\}$ and under-explores this path (i.e., $\ell_i^*(t)\geq \ell^{(m)}(t)$) if $x_i(t)>\max\Big\{\frac{\alpha-\alpha_L}{\alpha_H-\alpha_L},\Bar{x}\Big\}$. Then by the monotonicity of $\ell_i^*(t)$ in $x_i(t)$, we can prove the existence of $x^{th}$ to satisfy
\begin{equation}\tag{26}
    \min\Big\{\frac{\alpha-\alpha_L}{\alpha_H-\alpha_L},\Bar{x}\Big\} \leq x^{th}\leq \max\Big\{\frac{\alpha-\alpha_L}{\alpha_H-\alpha_L},\Bar{x}\Big\}.
\end{equation}

In the following two subsections, we first suppose $\frac{\alpha-\alpha_L}{\alpha_H-\alpha_L}<\Bar{x}$ to prove (26). According to the definition of $\ell_i^*(t)$, we will prove $\ell_i^*(t)\leq \ell^{(m)}(t)$ for $x_i(t)=\frac{\alpha-\alpha_L}{\alpha_H-\alpha_L}$ by showing $Q_0^*(t+1)<Q_i^*(t+1)$, and then prove $\ell_i^*(t)\geq \ell^{(m)}(t)$ for $x_i(t)=\Bar{x}$ by showing $Q_0^*(t+1)>Q_i^*(t+1)$. 
For ease of exposition, we assume $\lambda=1$ such that $s(t)=1$ is always true. Note that $\lambda$ does not influence the conclusion in Proposition~2, as both policies hold the same arrival rate.

\subsection{Over-exploration Proof}
In this subsection, we want to prove that the myopic policy over-explores stochastic path $i$ under $x_i(t)\leq \min\Big\{\frac{\alpha-\alpha_L}{\alpha_H-\alpha_L},\Bar{x}\Big\}$, by showing $Q_0^*(t+1)<Q_i^*(t+1)$ according to (23). 

If $x_i(t)=\frac{\alpha-\alpha_L}{\alpha_H-\alpha_L}$ at current time, we have $\mathbb{E}[\alpha_i(t)|x_i(t)]=\alpha$ for risky path $i$. If the current user arrival chooses path $\pi(t)=i$, the travel latency $\mathbb{E}[\ell_i(t+1)|x_i(t),y_i(t)]$ at the next time slot of this path $i$ is
\begin{align}
    &\mathbb{E}\Big[\ell_i(t+1)\Big|\frac{\alpha-\alpha_L}{\alpha_H-\alpha_L},y_i(t)\Big]\notag\\=&\mathbb{E}[\alpha_i(t)|x_i(t)]\mathbb{E}[\ell_i(t)|x_i(t-1),y_i(t-1)]+\Delta \ell.\label{E_l_t+1}
\end{align}
While if $\pi(t)=0$, the travel latency $\ell_{0}(t+1)$ of this safe path 0 at the next time slot is
\begin{equation*}
    \ell_0(t+1)=\alpha\ell_0(t)+\Delta \ell,
\end{equation*}
which equals $\mathbb{E}\big[\ell_i(t+1)\big|\frac{\alpha-\alpha_L}{\alpha_H-\alpha_L},y_i(t)\big]$ in (\ref{E_l_t+1}) under the condition $x_i(t)=\frac{\alpha-\alpha_L}{\alpha_H-\alpha_L}$. While the expected travel latency on any risky path $i$ becomes shorter than that under $\pi(t)=i$, as there is no user exploring to add congestion on it. Therefore, the cost-to-go satisfies $Q_0^*(t+1)<Q_i^*(t+1)$ in (23).

Next, we consider the case with $x_i(t)<\Bar{x}$. According to (10), the expected hazard belief at the next time slot satisfies $\mathbb{E}[x_i(t+1)]>x_i(t)$ by (10). According to (\ref{E_l_t+1}), the future expected travel latency on path $i$ will be longer than the travel latency on path 0 under $\pi(t)=0$. While for any other risky path $j$ with $j\neq i$, the dynamics of their expected travel latencies at future time slots are not dependent on current $\pi(t)=0$ or $\pi(t)=i$. As a result, the cost-to-go also satisfies $Q_0^*(t+1)<Q_i^*(t+1)$ in (23). 

In summary, users under the myopic policy over-explore path~$i$ given $x_i(t)\leq \frac{\alpha-\alpha_L}{\alpha_H-\alpha_L}$.

\subsection{Under-exploration Proof}
In this subsection, we prove that the myopic policy under-explores stochastic path $i$ under $x_i(t)\geq \max\Big\{\frac{\alpha-\alpha_L}{\alpha_H-\alpha_L},\Bar{x}\Big\}$, by showing $Q_0^*(t+1)>Q_i^*(t+1)$ according to (23).

If $x_i(t)=\Bar{x}$ at current time, we will prove that $Q_0^*(t+1)> Q_i^*(t+1)$ is always true for any time horizon $T>1$ using mathematical induction. 

If $T=1$ and $\ell_0(t)<\mathbb{E}[\ell_i(t)|x_i(t-1),y_i(t-1)]$, the myopic policy must choose path 0 but the socially optimal policy may choose $i$. We have 
\[
\begin{aligned}
    Q_i^*(t+1)&\leq \mathbb{E}[\ell_0(t+1)|\pi(t)=i]\\&<\min_j \mathbb{E}[\ell_j(t+1)|\pi(t)=0]=Q^*_0(t+1),
\end{aligned}
\]
because $\mathbb{E}[\alpha_i(t)|x_i(t)=\Bar{x}]>\alpha$. 

We suppose $Q_0^*(t+1)> Q_i^*(t+1)$ is also true for any time horizon $2\leq T\leq n$. Then we prove it is still true for time horizon $T=n+1$. 

Let $\pi_0^*(\tau)$ and $\pi_i^*(\tau)$ denote the two optimal policies for $Q_0^*(t+1)$ and $Q_i^*(t+1)$ after $t$, where $\tau\in\{t+1,\cdots,T+1\}$. If $\pi_0^*(\tau)=i$ again after time $t$, we let $\pi_i^*(\tau)=0$ to reach a lower cost $\mathbb{E}[\ell_{0}(\tau)|\pi(t)=i]< \mathbb{E}[\ell_i(\tau)|\pi(t)=0]$, due to the fact that $x_i(\tau)=\Bar{x}>\alpha$ for any $\tau$. Then for any other time slot $\tau\in\{t+1,\cdots,n+1\}$, we let $\pi_i^*(\tau)=\pi_0^*(\tau)$ to make $\mathbb{E}[\ell_{\pi_i^*(\tau)}(\tau)|\pi(t)=i]\leq \mathbb{E}[\ell_{\pi_{0}^*(\tau)}(\tau)|\pi(t)=0]$. After summing up all these costs, we can obtain $Q_0^*(t+1)> Q_i^*(t+1)$. In summary, users under the myopic policy under-explore path~$i$ given $x_i(t)\geq \Bar{x}$.

If $\frac{\alpha-\alpha_L}{\alpha_H-\alpha_L}>\Bar{x}$, we can use the same method to prove $\ell_i^*(t)\geq \ell^{(m)}(t)$ for $x_i(t)=\Bar{x}$ by showing $Q_0^*(t+1)>Q_i^*(t+1)$, and then prove $\ell_i^*(t)\leq \ell^{(m)}(t)$ for $x_i(t)=\frac{\alpha-\alpha_L}{\alpha_H-\alpha_L}$ by showing $Q_0^*(t+1)<Q_i^*(t+1)$. This completes the proof of the existence of $x^{th}$ in (26).

\section{Proof of Proposition 3}\label{proof:prop3}
If discount factor $\rho=0$, the optimal policy is the same as the myopic policy to only focus on the current cost. Then $\text{PoA}^{(m)}=\frac{1}{1-\rho}=1$ and the proposition holds. We next suppose that $\rho\in(0,1)$ to show that $\text{PoA}^{(m)}\geq \frac{1}{1-\rho}$.

We consider the simplest two-path transportation network with $N=1$ to purposely choose the proper parameters to create the worst case. Let $\ell_0(t)=\frac{\Delta \ell}{1-\alpha^{\frac{1}{\lambda}}}, x_1(t)=\Bar{x}, \mathbb{E}[\alpha_1(t)|\Bar{x}]=1,\ell_0(t)=\mathbb{E}[\ell_1(t)|\Bar{x},\emptyset],\alpha_L=0,q_{LL}\rightarrow 1$ and $\frac{\ell_0(t)}{\Delta \ell}\rightarrow\infty$. For other parameters, e.g., $q_{HH},\alpha_H$ and so forth, we can find proper values to satisfy the above constraints. Next, we will calculate the social costs under the myopic policy and socially optimal policy, respectively.

\subsection{Social Cost under Myopic Policy}
We first calculate the social cost under the myopic policy. As $\ell_0(t)=\ell_1(t)$, the current user will myopically choose safe path 0. 
Given $\ell_0(t)=\frac{\Delta\ell}{1-\alpha^{\frac{1}{\lambda}}}$, the expected travel latency $\ell_0(t+1)$ at the next time slot is
\begin{align*}
    \ell_0(t+1)&=\alpha^{\frac{1}{\lambda}}\ell_0(t)+\Delta\ell\\
    &=\frac{\Delta \ell}{1-\alpha^{\frac{1}{\lambda}}}=\ell_0(t).
\end{align*}
This means even though users keep arriving and choosing this safe path 0, the travel latency on this path is always no longer than $\frac{\Delta \ell}{1-\alpha^{\frac{1}{\lambda}}}=\ell_1(t)$.

As $\mathbb{E}[\alpha_1(t)|\Bar{x}]=1$ and $\ell_0(t)=\mathbb{E}[\ell_1(t)|\Bar{x},\emptyset]$ given $x_1(t-1)=\Bar{x}$, $y_1(t-1)=\emptyset$, we can obtain travel latency $\mathbb{E}[\ell_1(t+1)|\Bar{x},\emptyset]$ without exploration as
\begin{align*}
    \mathbb{E}[\ell_1(t+1)|\Bar{x},\emptyset]&=\mathbb{E}[\alpha_1(t)|x_1(t)=\Bar{x}] \mathbb{E}[\ell_1(t)|\Bar{x},\emptyset]\\
    &=\mathbb{E}[\ell_1(t)|\Bar{x},\emptyset].
\end{align*}
Since $x_1(t)=\Bar{x}$, the expectation $\mathbb{E}[x_1(t+1)]=\Bar{x}$ is also the steady state. Hence, the travel latency on this path also keeps at $\mathbb{E}[\ell_1(t)|\Bar{x},\emptyset]$ without any user exploration.

Under these parameters, myopic users keep choosing safe path 0 and will never explore risky path 1. Under user's arrival rate $\lambda$, we calculate the corresponding social cost as:
\begin{align*}
    C^{(m)}(\mathbf{L}(t),\mathbf{x}(t),s(t))&=\rho^{\frac{1}{\lambda}}\ell_0(t)+\rho^{\frac{2}{\lambda}}\ell_0(t) +\cdots\\
    &=\frac{\rho^{\frac{1}{\lambda}} \ell_0(t)}{1-\rho^{\frac{1}{\lambda}}}.
\end{align*}

\subsection{Minimum Social Cost}
Next, we will calculate the social cost under the socially optimal policy. We let the current user explore path 1 to exploit possible low-hazard state $\alpha_L=0$. As $q_{LL}\rightarrow 1$, $q_{HH}\rightarrow 1$, $p_H\rightarrow1$ and $p_L\rightarrow 0$, we obtain $P(y_1(t)=1)\rightarrow 0$ and $P(y_1(t)=0)\rightarrow 1$ as the system has been running for a long time. Then given user's arrival rate $\lambda$, we have
\begin{align*}
    C^*(\mathbf{L}(t),\mathbf{x}(t),s(t))\leq& \rho^{\frac{1}{\lambda}} \mathbb{E}[\ell_1(t)|\Bar{x},\emptyset]+\\&\text{Pr}(y_1(t)=1)\sum_{k=1}^\infty\rho^{\frac{k}{\lambda}}\ell_0(t+k)+\\&\text{Pr}(y_1(t)=0)\sum_{m=1}^\infty\rho^{\frac{m}{\lambda}} \mathbb{E}[\ell_1(t+m)|0,0],
\end{align*}
where $x_1(t+m)\rightarrow 0$ and $y_1(t+m)=0$ given $p_L\rightarrow 0$. The above inequality tells that if $y_1(t)=0$ with probability $\text{Pr}(y_1(t)=0)$, the socially optimal policy lets latter users keep exploiting the low travel latency on path 1, while if $y_1(t)=1$ with probability $\text{Pr}(y_1(t)=1)$, the socially optimal policy let the latter users go back to path 0 again. We further obtain 
\begin{align*}
    C^*(\mathbf{L}(t),\mathbf{x}(t),s(t))&\leq \rho^{\frac{1}{\lambda}}\mathbb{E}[\ell_1(t)|\Bar{x},\emptyset]+\frac{\rho^{\frac{2}{\lambda}}}{1-\rho^{\frac{1}{\lambda}}}\Delta\ell\\
    &=\ell_0(t)+\frac{\rho^{\frac{2}{\lambda}}}{1-\rho^{\frac{1}{\lambda}}}\Delta\ell.
\end{align*}
Finally, we can obtain 
\begin{align*}
    \text{PoA}^{(m)}&=\frac{C^{(m)}(\mathbf{L}(t),\mathbf{x}(t))}{C^*(\mathbf{L}(t),\mathbf{x}(t))}\\&\geq \frac{1}{1-\rho^{\frac{1}{\lambda}}},
\end{align*}
by letting $\frac{\Delta\ell}{\ell_0(t)}\rightarrow 0$.

\section{Proof of Proposition 4}\label{proof:prop4}
To tell the huge PoA result and selfish users' deviation from the optimal routing recommendations $\pi^*(t)$, we consider the simplest two-path transportation network. Initially, the safe path 0 has $\ell_0(t=0)=0$ with $\alpha\rightarrow 1$ and the risky path~1 has an arbitrarily travel latency $\ell_1(0)$. We let $\Bar{x}=0$ and $\mathbb{E}[\alpha_1(t)|\Bar{x}]=0$ by setting $q_{LL}=1$ and $q_{HH}\neq 1$. Then selfish users always choose $\pi^{\emptyset}(t)=1$ from $t=0$. We calculate the social cost
\begin{align*}
    C^{\emptyset}(\ell_0(0),\ell_1(0),\Bar{x},s(t))&=\rho^{\frac{1}{\lambda}}\ell_1(0)+\sum_{j=1}^\infty \rho^{\frac{j}{\lambda}} \Delta \ell\\
    &=\rho^{\frac{1}{\lambda}}\ell_1(0)+\frac{\rho^{\frac{2}{\lambda}} \Delta \ell}{1-\rho^{\frac{1}{\lambda}}},
\end{align*}
based on the fact that $\mathbb{E}[\alpha_1(t)|\Bar{x}]=0$ and $\mathbb{E}[\ell_1(t)|\Bar{x},y_1(t-1)]=\Delta\ell$ for any $t\geq 1$. 

However, the socially optimal policy lets users path 0 to exploit the small travel latency $\ell_0(0)$. In this case, we have
\begin{align*}
    C^*(\ell_0(0),\ell_1(0),\Bar{x},s(t))&=\rho^{\frac{1}{\lambda}}\ell_0(0)+\sum_{j=2}^\infty\rho^{\frac{j}{\lambda}}\Delta\ell\\
    &=\frac{\rho^{\frac{2}{\lambda}}\Delta\ell}{1-\rho^{\frac{1}{\lambda}}},
\end{align*}
where $\ell_0(0)=0$, $\mathbb{E}[\ell_1(1)|\Bar{x},\emptyset]=0$ and $\mathbb{E}[\ell_1(t)|\Bar{x},y(t-1)]=\Delta\ell$ for ant $t\geq 2$.

In this case, we obtain 
\begin{align*}
    \text{PoA}^{\emptyset}&=\max_{\ell_0(0),\ell_1(0),\Bar{x},\rho,\Delta\ell} \frac{C^{\emptyset}(\ell_0(0),\ell_1(0),\Bar{x},s(t))}{ C^*(\ell_0(0),\ell_1(0),\Bar{x},s(t))}\\&=\frac{(1-\rho^{\frac{1}{\lambda}})\ell_1(0)}{\rho^{\frac{2}{\lambda}}\Delta\ell}+\rho^{\frac{1}{\lambda}}\\ &=\infty,
\end{align*}
where we let $\frac{(1-\rho^{\frac{1}{\lambda}})\ell_1(0)}{\rho^{\frac{2}{\lambda}}\Delta\ell}\rightarrow\infty$.

Next, we analyze that even if the mechanism provides the optimal recommendation $\pi^*(t)$, the PoA is still infinite. Given the same belief on the initial travel latency $\ell_0(0)$ and $\ell_1(0)$, the selfish user will always choose the path $i$ because
\begin{align*}
    \text{Pr}\Big(\mathbb{E}[\ell_0(t)]>\mathbb{E}[\ell_i(t)|\Bar{x},y_i(t-1)]\Big)\rightarrow 1
\end{align*}
given $\mathbb{E}[\alpha_i(t)|\Bar{x}]=0$ and $\alpha\rightarrow 1$. Thus, the information hiding mechanism may not work given $\Bar{x}<\frac{\alpha-\alpha_L}{\alpha_H-\alpha_L}$, and it still makes $\text{PoA}^{\emptyset}=\infty$.

\section{Proof of Theorem 1}\label{proof:thm1}
We first prove that selfish users will follow our mechanism's optimal routing recommendations if $\Bar{x}>\frac{\alpha-\alpha_L}{\alpha_H-\alpha_L}$. After that, we prove that the worst-case PoA is reduced to $\frac{1}{1-\frac{\rho^{\frac{1}{\lambda}}}{2}}$ under $\Bar{x}<\frac{\alpha-\alpha_L}{\alpha_H-\alpha_L}$.

\subsection{Proof of SID Mechanism's Efficiency}
Note that if $\mathbb{E}[\alpha_i(t)|\Bar{x}]>1$, the expected travel latency on path $i$ keeps increasing exponentially in $\mathbb{E}[\alpha_i(t)|\Bar{x}]$. Given the system has been running for a long time, the socially optimal policy will never choose this path, either. Thus, we only consider the more practical case $\mathbb{E}[\alpha_i(t)|\Bar{x}]\leq 1$ in the following. 

Lacking any historical information on the hazard belief and assuming that the mechanism has already operated for a long time, the current user's best estimate of the travel latency of path $i$ is the stationary distribution $P^*(\Bar{\ell}_i(t))$ of $\Bar{\ell}_i(t)$ under optimal policy $\pi^*(t)$. We do not need to obtain $P^*(\Bar{\ell}_i(t))$ but can use it to estimate the long-run average un-discounted travel latency $\mu^*$ of all path:
\begin{align*}
    \mu^*&=\int_{A}\Bar{\ell}_i(t)dP^*(\Bar{\ell}_i(t))+\int_{B}\Bar{\ell_0}dP^*(\Bar{\ell}_i(t))\\
    &\leq \Bar{\ell}_0,
\end{align*}
where socially optimal policy chooses path $\pi^*(t)=i$ when $\Bar{\ell}_i(t)$ is in region A and chooses path $\pi^*(t)=0$ when $\Bar{\ell}_i(t)$ is in region B.

If the platform's recommendation is $\pi^*(t)=0$ for the current user, he will follow this recommendation to path 0. Otherwise, he will calculate
\begin{align*}
    &\mathbb{E}_{P^*(\Bar{\ell}_i(t))}[\Bar{\ell}_i(t)|\Bar{\ell}_i(t)\in A]\\=&\frac{\int_{A}\Bar{\ell}_i(t)dP^*(\Bar{\ell}_i(t))}{\int_{A}dP^*(\Bar{\ell}_i(t))}\\ =&\frac{\mu^*-\int_{B}\Bar{\ell_0}dP^*(\Bar{\ell}_i(t))}{\int_{A}dP^*(\Bar{\ell}_i(t))}\\ \leq&\frac{\Bar{\ell}_0-\int_{B}\Bar{\ell_0}dP^*(\Bar{\ell}_i(t))}{\int_{A}dP^*(\Bar{\ell}_i(t))}=\Bar{\ell}_0(t).
\end{align*}
Thus, each user will follow the optimal recommendation to choose path $\pi^*(t)=i$ given $\Bar{x}>\frac{\alpha-\alpha_L}{\alpha_H-\alpha_L}$. This is because the former exploration to risky path $i$ may learn a $\alpha_L$ there to greatly reduce $\mathbb{E}[\ell_i(t)|x_i(t-1),y_i(t-1)=0]$.

\subsection{Proof of $\text{PoA}^{(\text{SID})}\leq \frac{1}{1-\frac{\rho^{\frac{1}{\lambda}}}{2}}$}
Based on the former analysis, we can further prove $\text{PoA}^{(\text{SID})}\leq \frac{1}{1-\frac{\rho^{\frac{1}{\lambda}}}{2}}$, which only happens when $\Bar{x}<\frac{\alpha-\alpha_L}{\alpha_H-\alpha_L}$. With the information disclosure $\mathbf{L}(t)$, selfish users will deviate to follow myopic policy $\pi^{(m)}(t)$. We consider the maximum over-exploration in the simplest two-path network to show the bounded $\text{PoA}^{(\text{SID})}$.

Let $\ell_0(0)=\ell_1(0)-\varepsilon$ for path 0 with $\alpha\rightarrow 1$ to keep the travel latency on path 0 unchanged without user routing, where $\varepsilon$ is positive infinitesimal. We set $\ell_1(0)=\frac{\Delta \ell}{1-\mathbb{E}[\alpha_1(0)|\Bar{x}]^{\frac{1}{\lambda}}}$ for stochastic path 1 with $x_1(0)=\Bar{x}$, such that the expected travel latency at the next user arrival is
\begin{align*}
    \mathbb{E}_{y_1(0)}[\ell_1({\frac{1}{\lambda}})|\Bar{x},y_1(0)]&=\mathbb{E}[\alpha_1(0)|\Bar{x}]^{\frac{1}{\lambda}}\ell_1(0)+\Delta \ell \\ &=\frac{\Delta\ell}{1-\mathbb{E}[\alpha_1(0)|\Bar{x}]^{\frac{1}{\lambda}}},
\end{align*}
which keeps as $\ell_1(0)$ all the time with users' continuous explorations.
Then users keep choosing this risky path 1 without exploiting path 0, and we can calculate the social cost
\begin{align*}
    C^{(\text{SID})}(\ell_0(0),\ell_1(0),\Bar{x},s(t))&=\sum_{j=0}^\infty\rho^{\frac{j}{\lambda}}\ell_1(0)
    \\&=\frac{\ell_1(0)}{1-\rho^{\frac{1}{\lambda}}}.
\end{align*}
However, the socially optimal policy makes $\pi^*(0)=0$ for the first user to bear the same travel latency on path 0. Then the expected travel latency on the next user arrival on path 1 is reduced to \[\mathbb{E}[\ell_1({\frac{1}{\lambda}})|\Bar{x},\emptyset]=\mathbb{E}[\alpha_1(0)|\Bar{x}]^{\frac{1}{\lambda}}\ell_1(0).\]
After the first exploitation on path 0, the travel latency on this path increases to $\ell_0(0)+\Delta\ell$. While the expected travel latency on path 1 is always less than $\ell_1(0)$ because
\begin{align*}
    \mathbb{E}[\ell_1(t+1)|\Bar{x},y_1(t)]=&\Bar{\alpha}_1\mathbb{E}[\ell_1(t)|\Bar{x},y_1(t-1)]+\Delta\ell\\ < &\mathbb{E}[\alpha_1(0)|\Bar{x}]\ell_1(0)+\Delta\ell=\ell_1(0)
\end{align*}
for any time $t\geq 1$. 

Note that if $\mathbb{E}[\alpha_1(t)|\Bar{x}]=0$, then the expected travel latency on path 1 is always $\Delta\ell$, and socially optimal policy will not choose path 0, either. If $\mathbb{E}[\alpha_1(t)|\Bar{x}]=1$, then the expected travel latencies on both paths are infinite with $\frac{\Delta\ell}{1-\mathbb{E}[\alpha_1(0)|\Bar{x}]^{\frac{1}{\lambda}}}\rightarrow\infty$, and the $\text{PoA}^{\emptyset}=1$. Hence, the worst-case does not happen when $\mathbb{E}[\alpha_1(t)|\Bar{x}]=0$ or $1$. We next derive the $\mathbb{E}[\alpha_1(0)|\Bar{x}]$ in the worst-case, which is denoted by $\Bar{\alpha}$. From the evolution of $\mathbb{E}[\ell_1(t+1)|\Bar{x},y_1(t)]$, we aim to minimize the first order derivative below
\begin{align*}
    &\frac{\partial (\mathbb{E}[\ell_1(t+1)|\Bar{x},y_1(t)]-\mathbb{E}[\ell_1(t)|\Bar{x},y_1(t-1)])}{\partial \mathbb{E}[\ell_1(t-2)|\Bar{x},y_1(t-2)]}\\
    =&\frac{\Delta\ell +(\Bar{\alpha}_1-1)(\Bar{\alpha}_1\mathbb{E}[\ell_1(t-2)|\Bar{x},y_1(t-2)]+\Delta \ell)}{\partial \mathbb{E}[\ell_1(t-2)|\Bar{x},y_1(t-2)]}\\
    =&\Bar{\alpha}_1^2-\Bar{\alpha}_1,
\end{align*}
where the minimum is reached at $\Bar{\alpha}_1=\frac{1}{2}$. Note that $\Bar{\alpha}_1=\frac{1}{2}$ well balances the expected travel latency and a single user's incurred latency $\Delta\ell$ to make the largest PoA. 

Then we can calculate the optimal social cost as
\begin{align*}
    C^*(\ell_0(0),\ell_1(0),\Bar{x},s(t))=& \rho^{\frac{1}{\lambda}}\ell_0(0)+\rho^{\frac{2}{\lambda}}\Bar{\alpha}_1\ell_1(0)\\&+\rho^{\frac{3}{\lambda}}(\Bar{\alpha}^2_1\ell_1(0)+\Delta \ell)+\cdots\\
    \geq &\rho^{\frac{1}{\lambda}}\ell_0(0)+ \sum_{j=1}^{\infty}\rho^{\frac{j}{\lambda}}\frac{\ell_1(0)}{2}\\=&\rho^{\frac{1}{\lambda}}\ell_0(0)+\frac{\rho^{\frac{2}{\lambda}} \ell_1(0)}{2-2\rho^{\frac{1}{\lambda}}}.
\end{align*}
Though socially optimal policy may still choose path 0 to reduce the expected latency for path 1 after a period, the caused average expected travel latency is still no less than $\frac{\ell_1(0)}{2}$. This is because the travel latency on path 0 increases to $\ell_0(0)+\Delta \ell$ after the first exploitation. 

Finally, we can obtain the worst-case PoA as
\begin{align*}
    \text{PoA}^{(\text{SID})}&=\max \frac{C^{(\text{SID})}(\mathbf{L}(t),\mathbf{x}(t),s(t))}{C^*(\mathbf{L}(t),\mathbf{x}(t),s(t))}\\&\leq \frac{\frac{\rho^{\frac{1}{\lambda}}\ell_1(0)}{1-\rho^{\frac{1}{\lambda}}}}{\rho^{\frac{1}{\lambda}}\ell_0(0)+\frac{\rho^{\frac{2}{\lambda}} \ell_1(0)}{2-2\rho^{\frac{1}{\lambda}}}}\\&=\frac{1}{1-\frac{\rho^{\frac{1}{\lambda}}}{2}}.
\end{align*}
This completes the proof.

\section{Proof of Corollary 1}
Since a portion $\phi\in(0,1)$ of users rely on a single source for information, these users will follow our SID mechanism to make their path decisions. 
The remaining $1-\phi$ portion can access transparent information and will therefore follow the myopic policy.
In this section, we prove that the PoA under the SID mechanism is bounded by $\text{PoA}^{(\text{SID})} \leq \max\Big\{\frac{1}{1-\frac{\rho^{\frac{1}{\lambda}}}{2}}, \frac{1}{1-(1-\phi)\rho^{\frac{1}{\lambda}}}\Big\}$ by analyzing both maximum-exploration and minimum-exploration cases.

First, in the maximum-exploration scenario (with the same parameters as Theorem 1), all users-whether they belong to the $\phi$ or $1-\phi$ portion-will follow the myopic policy to always choose risky path 1 under our SID mechanism. In this case, the caused PoA upper bound is $\frac{1}{1-\frac{\rho^{\frac{1}{\lambda}}}{2}}$, consistent with Theorem~1.

Next, in the minimum-exploration scenario (under the same parameter setting as Proposition 3), the myopic policy always opts for safe path 0. However, the SID mechanism and the socially optimal policy will recommend the first user to explore risky path 1 to learn $\alpha_1(t)=\alpha_L$. Given that $\alpha_L=0$, subsequent users will consistently choose this risky path to exploit $\ell_1(t)=\Delta \ell$.
Based on our proof of Proposition~3 in Appendix~E, we obtain the long-term minimum social cost as:
\begin{align*}
    C^*(\ell_0(0),\ell_1(0),\Bar{x},s(t))=\ell_0(0)+\frac{\rho^{\frac{2}{\lambda}}}{1-\rho^{\frac{1}{\lambda}}}\Delta\ell.
\end{align*}
Then we analyze the long-term social cost caused by our SID mechanism.
At each time $t$, there are two possible scenarios:
\begin{itemize}
    \item Probability $1-\phi$: The current user has access to transparent information and follows the myopic policy to choose safe path 0. Then in the next time slot, the travel latencies $\ell_0(t+1)$ and $\ell_1(t+1)$ remain unchanged from $\ell_0(t)$ and $\ell_1(t)$,  respectively, and the next user arrival will face the same two possible scenarios. 
    \item Probability $\phi$: The current user relies on a single information source and follows the optimal recommendation to explore the risky path (path 1). As a result, the next user arrival—whether or not they have transparent information—will exploit the low travel latency on the risky path.
\end{itemize}
Consequently, we calculate the caused long-term social cost according to the above two cases as:
\begin{align*}
    &C^{(\text{SID})}(\ell_0(0),\ell_1(0),\Bar{x},s(0))\\=&(1-\phi) (\ell_0(0)+\rho^{\frac{1}{\lambda}}C^{(\text{SID})}(\ell_0(0),\ell_1(0),\Bar{x},s(0)))\\&+\phi C^*(\ell_0(0),\ell_1(0),\Bar{x},s(t)).
\end{align*}
Solving the above equality, we obtain 
\begin{align*}
    &C^{(\text{SID})}(\ell_0(0),\ell_1(0),\Bar{x},s(0))\\\leq& \frac{(1-\phi)\ell_0(0)+\phi C^*(\ell_0(0),\ell_1(0),\Bar{x},s(t))}{1-(1-\phi)\rho^{\frac{1}{\lambda}}}.
\end{align*}
Then we obtain the PoA upper bound \begin{align*}
    \text{PoA}^{(\text{SID})}\leq \frac{1}{1-(1-\phi)\rho^{\frac{1}{\lambda}}}.
\end{align*}

Based on the above analysis, we obtain that $\text{PoA}^{(\text{SID})}\leq \max\Big\{\frac{1}{1-\frac{\rho^{\frac{1}{\lambda}}}{2}}, \frac{1}{1-(1-\phi)\rho^{\frac{1}{\lambda}}}\Big\}$.

\section{Proof of Proposition 5}\label{proof:prop5}
Under the linear path graph in Fig. 4(a), both myopic and socially optimal policies just repeat decision-making for $k+1$ times from O to D. Yet the dynamics of Markov chains in Fig.~4(b) change the prior PoA ratio in Proposition 3 to (32). Therefore, we still focus on a segment $j$ to prove PoA in (32).

We consider the simplest case with $N^j=1$ risky path for each segment $j$ between nodes $\text{D}_j$ and $\text{D}_{j+1}$. And we prove PoA in (32) using the same parameters as in Appendix~\ref{proof:prop3}, i.e., $\ell_{0^j}(t)=\frac{\Delta \ell}{1-\alpha^{\frac{1}{\lambda}}}, x_{1^j}(t)=\Bar{x}, \mathbb{E}[\alpha_{1^j}(t)|\Bar{x}]=1,\ell_{0^j}(t)=\mathbb{E}[\ell_{1^j}(t)|\Bar{x},\emptyset],\alpha_L=0,q_{LL}\rightarrow 1$ and $\frac{\ell_{0^j}(t)}{\Delta \ell}\rightarrow\infty$. Next, we similarly calculate the social costs under the myopic policy and the socially optimal policy, respectively.

According to our proof in Appendix \ref{proof:prop3}, users at each node $\text{O},\text{D}_1,\cdots, \text{D}_k$ will myopically choose safe path $0^j$ with travel latency $\ell_{0^j}(t)=\frac{\Delta \ell}{1-\alpha^{\frac{1}{\lambda}}}$ at each segment $\{\text{D}_j,\text{D}_{j+1}\}$. 
In consequence, we calculate the total social cost for this segment under an infinite time horizon as
\begin{align*}
    C_j^{(m)}(\mathbf{L}(t),\mathbf{x}(t),s(t))
    &=\frac{\rho^{\frac{1}{\lambda}}\ell_{0^j}(t)}{1-\rho^{\frac{1}{\lambda}}}.
\end{align*}

Next, we will calculate the social cost under the socially optimal policy. Similarly, the socially optimal policy lets the current user explore path~$1^j$ to find the possible $\alpha_L=0$. Then the social cost for later users is greatly reduced for this segment $\{\text{D}_j,\text{D}_{j+1}\}$. However, given the maximum variation $\sigma$ for the dynamic transition probability $q_{LL}(t)$, there is a maximum probability $\sigma$ of switching to the high-hazard state for path $1^j$. If users keep observing $\alpha_{1^j}(t)=\alpha_L$ with probability $1-\sigma$, the cost-to-go is always $\rho^{\frac{1}{\lambda}}Q_{1^j}^*(t+\frac{1}{\lambda})$. While if any future user observes $\alpha_{1^j}(t)=\alpha_H$ with probability $\sigma$ there, the socially optimal policy will let future users switch back to safe path $0^j$, and the cost-to-go since then becomes $\rho^{\frac{1}{\lambda}}C_j^{(m)}(\mathbf{L}(t),\mathbf{x}(t)),s(t)\big)$. 
Therefore, the cost-to-go since the next user becomes
\begin{align*}
     Q_{1^j}^*(t+\frac{1}{\lambda})\leq& (1-\sigma)\big(\Delta \ell + \rho^{\frac{1}{\lambda}}Q_{1^j}^*(t+\frac{1}{\lambda})\big)+\\
    &\sigma \big(\ell_{0^j}(t+\frac{1}{\lambda})+ \rho^{\frac{1}{\lambda}}C_j^{(m)}(\mathbf{L}(t),\mathbf{x}(t),s(t))\big)\\
    \leq &\frac{(1-\sigma)\rho^{\frac{2}{\lambda}}}{1-(1-\sigma)\rho^{\frac{1}{\lambda}}}\Delta\ell+\frac{\sigma\rho^{\frac{2}{\lambda}}\ell_{0^j}(0)}{1-\sigma\rho^{\frac{1}{\lambda}}}.
\end{align*}

Given the current user's expected travel cost $\rho^{\frac{1}{\lambda}}\ell_{1^j}(0)$, we obtain the current social cost for the socially optimal policy as: 
\begin{align*}
    C^*(\mathbf{L}(t),\mathbf{x}(t),s(t))\leq& \rho^{\frac{1}{\lambda}}\ell_{1^j}(0)+\frac{(1-\sigma)\rho^{\frac{2}{\lambda}}}{1-(1-\sigma)\rho^{\frac{1}{\lambda}}}\Delta\ell\\&+\frac{\sigma\rho^{\frac{2}{\lambda}}\ell_{0^j}(0)}{1-\sigma\rho^{\frac{1}{\lambda}}}.
\end{align*}
Finally, we can obtain 
\begin{align*}
    \text{PoA}^{(m)}&=\frac{C^{(m)}(\mathbf{L}(t),\mathbf{x}(t),s(t))}{C^*(\mathbf{L}(t),\mathbf{x}(t),s(t))}\\&\geq \frac{1-\sigma\rho^{\frac{1}{\lambda}}}{1-\rho^{\frac{1}{\lambda}}},
\end{align*}
which proves (32) and completes the proof.

\section{Proof of Proposition 6}\label{proof:prop6}
We consider the same worst-case as Appendix \ref{proof:thm1}, where the myopic policy always explores risky path $1^j$ while the socially optimal policy may exploit safe path $0^j$ to reduce the travel latency on path $1^j$. In this, the travel latencies on safe path $0^j$ and risky path $1^j$ both keep at $\ell_0(t)=\mathbb{E}[\ell_1(t)|\Bar{x},y_1(t)]=\frac{\Delta\ell}{1-\mathbb{E}[\alpha_1(0)|\Bar{x}]^{\frac{1}{\lambda}}}$.
Then users keep choosing this risky path $1^j$ without exploitation to path $0^j$. To ensure the myopic policy's maximum exploration under the worst-case on risky path $1^j$, we make $\sigma\rightarrow 0$ to avoid state transition of this path.

Then given the average arrival rate $\lambda$, we can calculate the social cost under the myopic policy:
\begin{align*}
    C^{(m)}(\ell_{0^j}(0),\ell_{1^j}(0),\Bar{x})&=\sum_{j=1}^\infty\rho^{\frac{j}{\lambda}}\ell_{0^j}(0)
    \\&=\frac{\rho^{\frac{1}{\lambda}}\ell_{1^j}(0)}{1-\rho^{\frac{1}{\lambda}}}.
\end{align*}
However, the socially optimal policy makes $\pi^*(0)=0$ for the first arriving user to bear the similar travel latency on path $0^j$. Then the expected travel latency on the next time slot on path~$1^j$ is reduced to \[\mathbb{E}[\ell_{1^j}(\frac{1}{\lambda})|\Bar{x},\emptyset]=\mathbb{E}[\alpha_{1^j}(0)|\Bar{x}]^{\frac{1}{\lambda}}\ell_{1^j}(0).\]
After the first exploitation on path $0^j$, the travel latency on this path increases to $\ell_{0^j}(0)+\Delta\ell$. While the expected travel latency for the next user on path~$1^j$ is always less than $\ell_{1^j}(0)$:
\begin{align*}
    \mathbb{E}[\ell_{1^j}(t+\frac{1}{\lambda})|\Bar{x},y_{1^j}(t)]=&\Bar{\alpha}_{1^j}^{\frac{1}{\lambda}}\mathbb{E}[\ell_{1^j}(t)|\Bar{x},y_{1^j}(t-1)]+\Delta\ell\\ < &\mathbb{E}[\alpha_{1^j}(0)|\Bar{x}]^{\frac{1}{\lambda}}\ell_{1^j}(0)+\Delta\ell=\ell_{1^j}(0)
\end{align*}
for any time $t\geq 1$. 

According to Appendix \ref{proof:thm1}, the minimum social cost is reached at $\Bar{\alpha}_{1^j}=\frac{1}{2}$. Then we can similarly calculate the optimal social cost as
\begin{align*}
    C^*(\ell_{0^j}(0),\ell_{1^j}(0),\Bar{x})\geq &\rho^{\frac{1}{\lambda}}\ell_{0^j}(0)+\frac{\rho^{\frac{2}{\lambda}} \ell_{1^j}(0)}{2-2\rho^{\frac{1}{\lambda}}}.
\end{align*}
Finally, we can obtain the PoA as
\begin{align*}
    \text{PoA}^{(\text{SID})}&=\max \frac{C^{(m)}(\mathbf{L}(t),\mathbf{x}(t),s(t))}{C^*(\mathbf{L}(t),\mathbf{x}(t),s(t))}\\&\leq\frac{1}{1-\frac{\rho^{\frac{1}{\lambda}}}{2}}.
\end{align*}
This completes the proof.


\end{document}